\documentclass[conference]{IEEEtran}
\IEEEoverridecommandlockouts
\usepackage{cite}
\usepackage{amsmath,amssymb,amsfonts}
\usepackage{algorithmic}
\usepackage{graphicx}
\usepackage{textcomp}
\usepackage{xcolor}
\def\BibTeX{{\rm B\kern-.05em{\sc i\kern-.025em b}\kern-.08em
    T\kern-.1667em\lower.7ex\hbox{E}\kern-.125emX}}
\usepackage{graphicx}% Include figure files
\usepackage{xcolor}
\usepackage{dcolumn}% Align table columns on decimal point
\usepackage{bm}% bold math
\usepackage[utf8]{inputenc}
\usepackage[T1]{fontenc}
\usepackage{mathptmx}
\usepackage{etoolbox}
\usepackage{chemformula}
\usepackage{hyperref}
\usepackage{cleveref}
\usepackage{orcidlink}
\usepackage{makecell}

\begin{document}

\title{Classical and Quantum Frequency Combs for Satellite-based Clock Synchronization}

\author{
\IEEEauthorblockN{Ronakraj K. Gosalia\IEEEauthorrefmark{1}\IEEEauthorrefmark{4}, Ryan Aguinaldo\IEEEauthorrefmark{2}, Jonathan Green\IEEEauthorrefmark{2}, Holly Leopardi\IEEEauthorrefmark{3}, Peter Brereton\IEEEauthorrefmark{3} and Robert Malaney\IEEEauthorrefmark{1}}
\IEEEauthorblockA{\IEEEauthorrefmark{1}University of New South Wales, Sydney, NSW 2052, Australia.}
\IEEEauthorblockA{\IEEEauthorrefmark{2}Northrop Grumman Corporation, San Diego, CA 92128, USA.}
\IEEEauthorblockA{\IEEEauthorrefmark{3}NASA Goddard Space Flight Center, Greenbelt, MD 20771, USA.}
\IEEEauthorblockA{\textit{\IEEEauthorrefmark{4}Electronic mail: \href{mailto:r.gosalia@unsw.edu.au}{r.gosalia@unsw.edu.au}}}
}

\maketitle
\thispagestyle{plain}
\pagestyle{plain}

\begin{abstract}
The next generation of space-based networks will contain optical clocks embedded within satellites. To fully realize the  capabilities of such clocks, high-precision clock synchronization across the networks will be necessary. Current experiments have shown the potential for classical frequency combs to synchronize remote optical clocks over free-space. However, these classical combs are restricted in precision to the standard quantum limit. Quantum frequency combs, however, which exhibit quantum properties such as squeezing and entanglement, provide pathways for going beyond the standard quantum limit. Here, we present our perspective on the prospects for practical clock synchronization in space using both classical and quantum frequency combs. We detail the current outcomes achievable with a classical frequency comb approach to synchronization, before quantifying the potential outcomes offered by quantum frequency combs. Challenges to be overcome in deploying frequency combs in space are presented,  and the implications of almost-perfect synchronization for future space-based applications and experiments discussed. 
\end{abstract}

\section{Introduction}

The advent of optical clocks has enabled an unprecedented level of stability, accuracy, and precision in timekeeping\cite{Ludlow2008,Poli2013a}, providing a viable solution for many applications. Indeed, optical clocks are expected to be a key enabler for next-generation metrology\cite{Katori2011}, astronomy\cite{Clivati2017}, geodesy\cite{Mehlstaubler2018a}, navigation\cite{Boldbaatar2023}, fundamental physics tests\cite{Derevianko2022}, and more\cite{Riehle2017}. However, the performance of optical clocks is strictly restricted in a network by the performance of the clock synchronization scheme used. 
%This presents a new challenge for modern-day clock synchronization schemes that, until recently, has been insurmountable. 
Clock synchronization across a network ensures that network operations across devices occur in the desired sequence. As the synchronization improves,
%approaches \textit{perfect} synchronization, 
the operations can proceed at faster rates, leading to higher network performance levels across a wide variety of network functions\cite{Johannessen2004}. Many of the applications and tests mentioned above are network based, and therefore also enhanced by improved synchronization.
Although many different schemes have previously been proposed to minimize error during clock synchronization\cite{Krummacker2020,Gore2020,Seijo2022}, it is now emerging that 
optical-based protocols based on frequency combs (that interconnect optical clocks)
are enabling precision of clock synchronization at the standard quantum limit\cite{Giorgetta2013,Deschenes2016a,Ellis2021a,Caldwell2022,Caldwell2022b,Caldwell2023} (SQL) and beyond\cite{DeBurgh2005,Komar2014,Lamine2008,Wang2018,Gosalia2022,Gosalia2023}.
%At this level of performance, near-perfect synchronization could be achieved that would enable a new generation of communication, navigation and sensing protocols.
%Current commercially operating clock synchronization schemes are based on microwave signals, such as the global positioning system (GPS), and have inadequate precision for synchronizing optical clocks. Instead, the field is transitioning to optical frequency-based schemes for increased precision and several different techniques are being intensively studied.
%, a prime candidate is yet to be found. 

Beyond frequency combs, other optical-based synchronization strategies currently being investigated include techniques focusing on continuous-wave laser\cite{Djerroud2010a,Samain2015,Rovera2016,Gozzard2018,Shen2022}, chirped frequency\cite{Raupach2014a}, quantum correlated photons\cite{Jozsa2000,Ilo-Okeke2018,Ilo-Okeke2020,Dai2020,Spiess2023a,Lafler2023,Haldar2023} and pulsed single photons\cite{Spiess2024a}, to name a few. However, notwithstanding the merits of these other strategies, in  this work we focus on the progress of classical\cite{Giorgetta2013,Deschenes2016a,Ellis2021a,Caldwell2022,Caldwell2022b,Caldwell2023} and quantum\cite{DeBurgh2005,Komar2014,Lamine2008,Wang2018,Gosalia2022,Gosalia2023} frequency combs as, in our view, they show the most promise for practical synchronization improvement over free-space links\cite{Caldwell2023} especially  links between satellites\cite{Gosalia2022,Gosalia2023}, and ground-satellite configurations\cite{Giorgetta2013,Ellis2021a,Kang2019}. 

For our purposes, the distinction made  between ``classical'' and ``quantum'' frequency combs will lie within whether the performance (defined by the  precision in estimating the timing  during clock synchronization) is limited by the SQL or the Heisenberg limit (HL). As discussed more later, the HL has a better scaling with the system resource, $n$, used; with, in the context of synchronization, timing accuracy improved by a factor of $\sqrt n$ relative to the SQL scaling. The HL is known to be  the fundamental scaling performance achievable by any system --- and requires the introduction of quantum processes. The SQL is the best-case performance of a system based on classical properties of light, such as a classical frequency comb. The quantum frequency combs can instead approach the more fundamental HL. In this paper, the term quantum frequency comb will refer to a pulsed-laser system that exhibits either quadrature-squeezing or quadrature-entanglement (detailed further in \cref{sec:quantum-combs}).
%and is in practice generated by converting a classical frequency comb and a non-linear optical process (detailed further in \cref{sec:quantum-combs}). 
Other types of quantum frequency combs based on single photons\cite{Kues2019} (which are also called ``quantum optical microcombs'') are beyond the scope of this work.

\begin{figure}
    \centering
    \includegraphics[width=\linewidth]{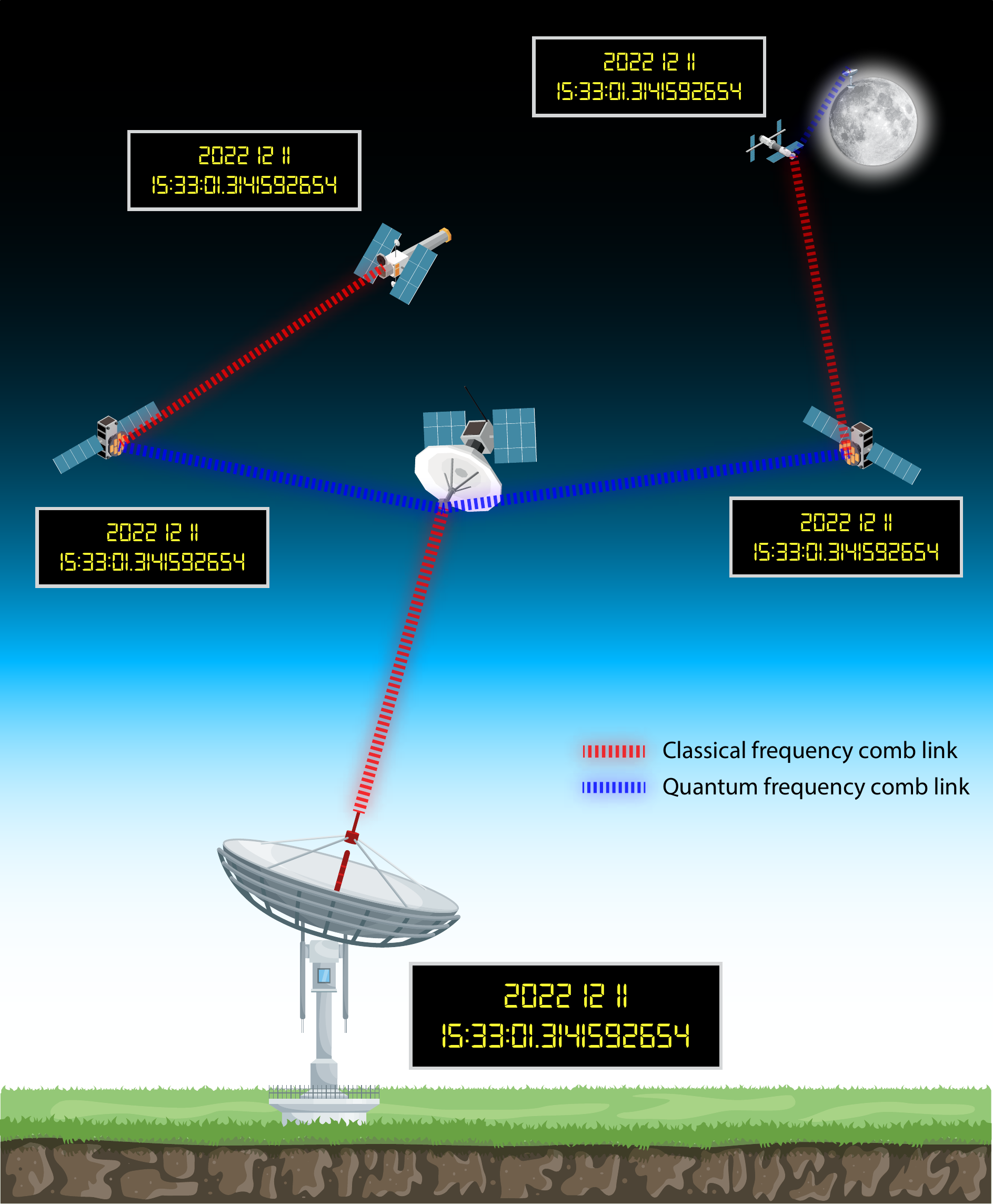}
    \caption{Our perspective on the next-generation network of highly-synchronized satellite-based clocks which will be the backbone infrastructure for many applications including communications, navigation and sensing. A mixture of classical and quantum frequency comb-based links are expected based on current research. The quantum approach provides, in principle, resource efficiencies for high-precision clock synchronization which are important in low SWaP (e.g. satellite) scenarios (as discussed in \cref{sec:background}). However, at present, the quantum approach is limited to short-range inter-satellite links where the detrimental impact of loss and noise can be contained (as detailed in \cref{sec:quantum-combs}). On the contrary, over long-range links such as ground-satellite, the classical approach may be preferred due to implementation simplicity, but this would be at the cost of a higher SWaP demand (see \cref{sec:classical}). An all-quantum approach would maximize the resource efficiencies across the entire network, but many implementation challenges currently exist which are open directions of research.}
    \label{fig:figure1}
\end{figure}

It is our view that over size, weight and power (SWaP) constrained links, the quantum frequency comb can provide an efficient solution for next generation satellite networks that require fundamental scaling performance. We envisage a scenario where both classical and quantum frequency combs co-exist throughout the network, the latter only being used when absolutely required on certain links.
The architecture in \cref{fig:figure1} summarizes this future perspective.
The delivery, and performance of the classical and quantum frequency combs that form the synchronization links of this space-based architecture, forms the focus of our paper.

%We emphasize that the quantum-based schemes of interest to us have a different objective to other quantum schemes based on correlated photons\cite{Jozsa2000,Ilo-Okeke2018,Ilo-Okeke2020,Dai2020,Spiess2023a,Lafler2023,Haldar2023}. The latter schemes mainly focus on the security advantages afforded by quantum properties and, consequently, have precision limitations\cite{Haldar2023}. Instead, our focus is on maximizing the precision per unit resource consumed (i.e. per photon). 

%It is our view that over size, weight and power (SWaP) constrained links, the quantum frequency comb can provide an efficient solution.
%Finally, for the purpose of this work, our usage of the term ``quantum frequency comb'' needs to be ascertained. %Quantum frequency combs can take the form of either single (and bi-) photon, or multi-photon pulses; here, we focus on the multi-photon case which requires the continuous-variable treatment of light since, as shown in \cref{fig:figure2} of \cref{sec:background}, more signal photons lead to a greater potential for a precision-to-resource advantage.

This rest of this paper is organized as follows. In \cref{sec:background}, we provide contextual details on the recent progress in timekeeping and clock synchronization, and compare the fundamental precision advantage of quantum-enhanced clock synchronization. We provide details of classical frequency combs in \cref{sec:classical}, including the operating principles, a comparison of recent satellite-based experiments, and highlight the key challenges ahead. In \cref{sec:quantum-combs}, we highlight the theory, recent progress, and challenges of quantum frequency combs, and  \cref{sec:perspective} includes our outlook on the field. Finally, \cref{sec:conclusion} presents our conclusions.

\section{\label{sec:background}Background}
In this section, we briefly discuss the recent progress in optical clocks and synchronization. Further, trade-offs between classical and quantum clock synchronization schemes are discussed. 
%For brevity, we use the term ``uncertainty'' to refer to the fractional frequency instability and use this term as a metric for comparing different clock synchronization protocols. 
As mentioned earlier, satellite-based optical clocks would enable a new frontier for next-generation communications, navigation and sensing applications. These applications will all require a network of clocks. This network would also need time-transfer capabilities at precision levels that match, and ideally exceed, the timekeeping precision of the clocks. Hence, as the field of optical clocks progresses, so to must the field of clock synchronization. %As a common metric for comparing between different clock technologies as well as between different clock synchronization techniques, we use the fractional frequency instability (FFI).
%and the timing deviation (TDEV).

\subsection{\label{sec:ffi-tdev}Fractional frequency instability and timing deviation}
%The metrics of main interest to us for comparing between the performance of different optical clocks and different clock synchronization schemes are the fractional frequency instability (FFI) and the timing deviation (TDEV). 
In the literature, the term fractional frequency instability (FFI) is also referred to as the ``fractional frequency uncertainty'', the ``Allan deviation'', the ``systematic uncertainty'' or simply as the ``uncertainty''. Technically, the FFI is the average two-sample difference of the average fractional frequency and is a useful metric for comparing between different clocks and clock synchronization techniques\cite{IEEE2022}. The fractional frequency is a ratio of the difference between the oscillating frequency and the nominal frequency divided by the nominal frequency. In the context of clocks, the FFI is a measure of the frequency stability of a clock, and a smaller value of FFI represents a higher clock stability and time-keeping precision. Across the literature, however, different methods are used to calculate the FFI and sometimes the method used is not specified, which makes comparisons challenging. Here, we begin by detailing the three most popular definitions for the FFI as outlined in IEEE 1139-2022\cite{IEEE2022} and by Riley et al.\cite{Riley2008}. %Additionally, whether the ``Allan deviation'' specified in papers refers to the standard or modified versions are not always made clear. 

An instantaneous signal produced by the oscillations in a clock can be described as a time-dependent amplitude, $S(t)$, given by
\begin{gather}
    S(t)=S_0\cos(2\pi\nu_0t+\phi(t)),%\equiv S_0\cos(2\pi\nu_0[t+x(t)]),
\end{gather}
where $S_0$ is the peak amplitude, $\nu_0$ is the nominal oscillating frequency of the clock, %${\phi(t):=2\pi\nu_0x(t)}$ 
and $\phi(t)$ is the instantaneous phase of the clock. %, and $x(t)$ is the instantaneous time deviation from a nominal time. 
When characterizing clocks and clock synchronization techniques, we are concerned with the stability of two parameters: the frequency and the phase. Instabilities in the frequency are calculated via the instantaneous fractional frequency, $y(t)$, which is given by
\begin{gather}
    y(t) = \frac{\nu(t)-\nu_0}{\nu_0}\equiv\frac{1}{2\pi\nu_0}\frac{d\phi(t)}{dt},
    \label{eq:yt}
\end{gather}
where $\nu(t)$ is the instantaneous frequency of the clock. Instabilities in the phase are calculated via the instantaneous timing deviation, $x(t)$, that is given by
\begin{gather}
    x(t) = \frac{\phi(t)-\phi_0}{2\pi\nu_0}, 
    \label{eq:xt}
\end{gather}
where $\phi_0$ is the nominal phase (often assumed $\phi_0=0$). We also note that $y(t)$ and $x(t)$ are related to each other via
\begin{gather}
    y(t) = \frac{dx(t)}{dt}.
    \label{eq:yt-xt}
\end{gather}
An experimenter has the choice regarding whether to measure $y(t)$, $x(t)$ or both.
In practice, samples of $y(t)$ or $x(t)$ are collected at discrete time intervals separated by a fixed sample period $\tau_0$, giving a collection of samples $[\bar{y}_1, \bar{y}_2,\cdots]$ and $[\bar{x}_1,\bar{x}_2,\cdots]$, respectively. Each sample can be described as an average quantity over $\tau_0$, namely
\begin{gather}
    \bar{y}_k = \frac{1}{\tau_0}\int_{t_k}^{t_{k+1}}y(t)dt\quad\text{and}\quad \bar{x}_k=\frac{1}{\tau_0}\int_{t_k}^{t_{k+1}}x(t)dt,
    \label{eq:yk}
\end{gather}
with $t_k=(k-1)\tau_0$ and $k\in[1,2,\ldots]$ is the sample index. Note, also the discrete version of \cref{eq:yt-xt} used in practice, namely
\begin{gather}
    \bar{y}_k = \frac{\bar{x}_{k+1}-\bar{x}_{k}}{\tau_0}.
\end{gather}
%Note, in \cref{eq:yt-xt} we have also given the relationship between the instantaneous fractional frequency and the instantaneous time deviation. 
%In practice, $y(t)$ is averaged over a finite integration time-window which we denote throughout this work as $\tau$. Several averages are collected in an experiment, and used to calculate the FFI. A single average of $y(t)$, denoted as $\bar{y}_k$ (where ${k\in[1,2,\ldots]}$ is an index), that is taken from time $t_k$ to $t_{k+1}$ with ${t_k=(k-1)\tau}$ is given by
%\begin{gather}
%    \overline{y}_k=\frac{1}{\tau}\int_{t_k}^{t_{k+1}}y(t)dt.%\equiv\frac{x_{k+1}-x_k}{\tau},
%    \label{eq:yk}
%\end{gather}
%Note, $\bar{y}_k$ is the average frequency deviation over the interval of length $\tau$ from $t_k$ to $t_{k+1}$.
%where $x_k:=x(t_0+k\tau_0)$ and $\tau_0$ is the fixed sampling period.
%Here, \cref{eq:yk} represents the average fractional frequency over $\tau$.
After $M\gg1$ samples are collected, the FFI is calculated. The first method for computing the FFI, which we denote $\text{FFI}^{(0)}$, calculates the average difference between adjacent samples, and is given by
\begin{gather}
    %\sigma_y(\tau) := \left[\frac{1}{2}\left\langle[\bar{y}_{k+1}-\bar{y}_{k}]^2\right\rangle\right]^{1/2},
    \nonumber \text{FFI}^{(0)} = \left[\frac{1}{2(M-1)}\sum_{k=1}^{M-1}\left(\overline{y}_{k+1}-\overline{y}_k\right)^2\right]^{1/2} \\
    = \left[\frac{1}{2(M-1)\tau_0^2}\sum_{k=1}^{M-1}\left(\bar{x}_{k+2}-2\bar{x}_{k+1}+\bar{x}_{k}\right)^2\right]^{1/2}.
    \label{eq:FFI}
\end{gather}
%In the form given in \cref{eq:FFI}, $y(t)$ is averaged over non-overlapping intervals of the integration time such that two consecutive time-windows do not share any samples of $y(t)$. %Also, \cref{eq:FFI} is the definition of the standard Allan deviation. 
The method in \cref{eq:FFI} has been largely superseded by a second method, denoted here as $\text{FFI}^{(1)}$, that uses the difference between non-adjacent samples to improve the confidence in the FFI estimate. The time-distance between the non-adjacent samples is an integer multiple of $\tau_0$ which we denote as $m\tau_0$ where $m\in\mathbb{Z}^+$. $\text{FFI}^{(1)}$ is given by
\begin{gather}
    \nonumber \text{FFI}^{(1)} = \left[\frac{1}{2m^2(M-2m+1)}\sum_{j=1}^{M-2m+1}\left\{\sum_{k=j}^{j+m-1}\left(\bar{y}_{k+m}-\bar{y}_k\right)\right\}^2\right]^{1/2} \\
    = \left[\frac{1}{2m^2(M-2m+1)\tau_0^2}\sum_{k=1}^{M-2m+1}\left(\bar{x}_{k+2m}-2\bar{x}_{k+m}+\bar{x}_k\right)^2\right]^{1/2}.
    \label{eq:FFI-overlap}
\end{gather}
Finally, a third method exists, denoted here as $\text{FFI}^{(2)}$, which has the advantage of acting as an algorithmic filter to help distinguishing between white noise and flicker phase noise as discussed later in this section. $\text{FFI}^{(2)}$ extends $\text{FFI}^{(1)}$ using the same time-distance between non-adjacent samples of $m\tau_0$, and is given by
\begin{gather}
    \nonumber \text{FFI}^{(2)} = \bigg[\frac{1}{2m^4(M-3m+2)}\times\\
    \nonumber \sum_{j=1}^{M-3m+2}\left\{\sum_{i=j}^{j+m-1}\left(\sum_{k=i}^{i+m-1}(\bar{y}_{k+m}-\bar{y}_k)\right)\right\}^2\bigg]^{1/2} \\
    \nonumber = \bigg[\frac{1}{2m^4(M-3m+2)\tau_0^2}\times\\
    \sum_{j=1}^{M-3m+2}\left(\sum_{k=j}^{m+j-1}\left(\bar{x}_{k+2m}-2\bar{x}_{k+m}+\bar{x}_k\right)\right)^{2}\bigg]^{1/2}.
    \label{eq:FFI-mod}
\end{gather}

\begin{figure}
    \centering
    \includegraphics[width=\linewidth]{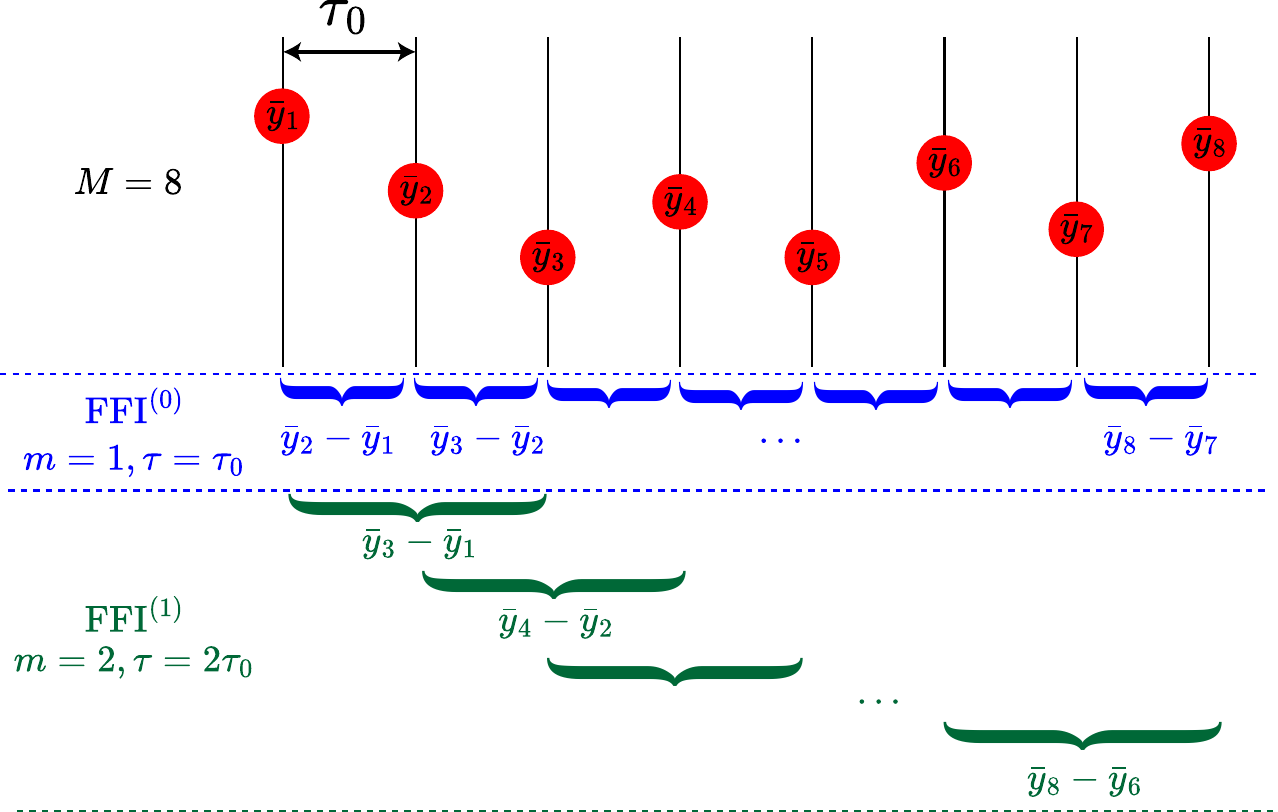}
    \caption{An example experiment with $M=8$ samples of $\bar{y}$ that are separated in time by the fixed sampling period $\tau_0$. Two different methods for computing the FFI are pictorially shown. The first, $\text{FFI}^{(0)}$, calculates the average difference between adjacent samples of $\bar{y}$. The second, $\text{FFI}^{(1)}$ uses non-adjacent samples --- here, as an example, we have fixed $m=2$ such that the time duration between the non-adjacent samples is $2\tau_0$. The integration time, $\tau$, is the duration of each sampling window; in the above example, with $\text{FFI}^{(0)}$, $\tau=\tau_0$, and with $\text{FFI}^{(1)}$, $\tau=2\tau_0$.}
    \label{fig:FFI0vsFFI1}
\end{figure}

When ${m=1}$ \cref{eq:FFI-overlap,eq:FFI,eq:FFI-mod} are all equal. Also, depending on the size of $m$, $\text{FFI}^{(0)}$ can be very different to $\text{FFI}^{(1)}$ and $\text{FFI}^{(2)}$ over the same collected samples.
To aid, in \cref{fig:FFI0vsFFI1} we have provided a visual of an example experiment where ${M=8}$ samples of $\bar{y}$ are collected. Here, the difference between $\text{FFI}^{(0)}$ and $\text{FFI}^{(1)}$ are shown when ${m=2}$. 
The integration time, denoted by $\tau$, is the duration of the sampling window in seconds. When $\text{FFI}^{(0)}$ is computed, ${\tau=\tau_0}$, and when $\text{FFI}^{(1)}$ or $\text{FFI}^{(2)}$ are computed ${\tau=m\tau_0}$. It is important that the FFI is always reported alongside $\tau$ due its dependence on $m$ and $\tau_0$.

Unfortunately, across the literature, there is an inconsistency in that some works report the $\text{FFI}^{(0)}$, while others report the $\text{FFI}^{(1)}$ or $\text{FFI}^{(2)}$. To make matters worse, some works do not make clear whether they are reporting the $\text{FFI}^{(0)}$, $\text{FFI}^{(1)}$ or $\text{FFI}^{(2)}$. In this paper, we will stipulate  whether $\text{FFI}^{(0)}$, $\text{FFI}^{(1)}$ or $\text{FFI}^{(2)}$ were used in a referenced work when it is clear. However, when a referenced work has failed to stipulate the specific form used we will term the reported value as simply the ``FFI''. As will be clear from the context in which it is used, we will also occasionally use the term ``FFI'' as a means to refer to either of $\text{FFI}^{(0)}$,
$\text{FFI}^{(1)}$ or $\text{FFI}^{(2)}$. In recent publications, $\text{FFI}^{(0)}$ is sometimes referred to as the ``Allan deviation'', while $\text{FFI}^{(1)}$ the ``overlapping Allan deviation'' and $\text{FFI}^{(2)}$ the ``modified Allan deviation''. %Most commonly, $\text{FFI}^{(0)}$ is referred to as the Allan deviation in the literature, while $\text{FFI}^{(1)}$ is the so-called ``modified'' Allan deviation.
The different definitions of FFI exist to help distinguish between various sources of noise in real experiments\cite{Allan1987} which, in the context of clocks and clock synchronization, are white noise, flicker noise, and random-walk\cite{Riley2008,IEEE2022}. These noise sources can be present in $\bar{x}$ and $\bar{y}$ measurements, giving rise to phase-modulation (PM) and frequency-modulation (FM), respectively. White noise is uncorrelated random noise which has a flat spectral density profile. Flicker noise has a spectral density given by $f^{-1}$, where $f$ is the frequency, and random-walk has a spectral density given by $f^{-2}$. When either of these noise sources dominate the measurement, the influence can be observed via the gradient of the log-log plot of FFI vs. $\tau$. The various gradient values for each noise source are summarized in \cref{tab:sigma-tau-plot}. For example, when white PM noise is dominant, the $\text{FFI}^{(0)},\text{FFI}^{(1)}\propto\tau^{-1}$ and $\text{FFI}^{(2)}\propto\tau^{-3/2}$. In a state-of-the-art clock synchronization experiment, white PM was the dominant noise source from $0\leq\tau\leq0.5$~s, and when $\tau>0.5$~s flicker PM noise was dominant\cite{Caldwell2023}. Note, the distinction between white PM and flicker PM can only be made with $\text{FFI}^{(2)}$, and this is the reason for the general preference for $\text{FFI}^{(2)}$ over the other definitions of FFI. Further, the noise sources also limit the maximum $\tau$ in a clock synchronization experiment --- the maximum $\tau$ is when the dominant noise source transitions from white PM to flicker PM noise\cite{IEEE2022}. FFI measurements beyond the maximum $\tau$ are not trusted.

%$\text{FFI}^{(0)}$, $\text{FFI}^{(1)}$ and $\text{FFI}^{(2)}$ all converges for the main sources of noise experienced in clocks\cite{Allan1987}.% --- where the usual standard deviation diverges\cite{Allan1987}. 
%Random uncorrelated noise sources are called white noise, %, given by
%\begin{gather}
%    \nonumber \sigma(M) = \left[\frac{1}{M-1}\sum_{k=1}^{M}(\bar{y}_k-\bar{y})^2\right]^{1/2},
%\end{gather}
%where $\bar{y}:=\left(\sum_{k=1}^{M}\bar{y}_k\right)/M$ is the average over all collected samples.

\begin{table*}[t]
    \centering
    \caption{The coefficient $\alpha$ of $\text{FFI}\propto\tau^{\alpha}$ for the most common sources of noise. Note, $\alpha$ is different for white PM and flicker PM when $\text{FFI}^{(2)}$ is used compared to $\text{FFI}^{(0)}$ and $\text{FFI}^{(1)}$, and for this reason $\text{FFI}^{(2)}$ is preferred to distinguish between these two sources of noise\cite{Riley2008}.}% Acronyms and symbols used are as follows, \textbf{Er}: Erbium, \textbf{MLL}: mode-locked laser, $f_r$: repetition frequency, $T_0$: pulse duration, $P_{in}$: input pump power and $P_{out}$: output seed power.}
    \label{tab:sigma-tau-plot}
    \begin{tabular}{c c c c c c}
        \Xhline{3\arrayrulewidth}
         FFI method & White PM & Flicker PM & White FM & Flicker FM & Random-walk FM \\%& Frequency drift \\
         \Xhline{2\arrayrulewidth}
         $\text{FFI}^{(0)}$ and $\text{FFI}^{(1)}$ & $-1$ & $-1$ & $-1/2$ & $0$ & $1/2$\\% & $1$ \\
         $\text{FFI}^{(2)}$ & $-3/2$ & $-1$ & $-1/2$ & $0$ & $1/2$\\% & $1$ \\
          \Xhline{3\arrayrulewidth}
    \end{tabular}
\end{table*}

Although the FFI provides a useful metric for comparing the stability of clocks, in clock synchronization we are interested in the timing error between clocks.
To this end we introduce the timing deviation
(TDEV) which quantifies the time stability of a clock. The TDEV is related to the FFI via,
\begin{gather}
    \text{TDEV}=\frac{\tau}{\sqrt{3}}\text{FFI}^{(2)},
    \label{eq:tdev}
\end{gather}
with units of seconds.
The timing error between clocks, denoted throughout this work by the standard deviation $\sigma_{\Delta t}$, is given by
\begin{gather}
    \sigma_{\Delta t} = \sigma_{excess} + \text{TDEV},
\end{gather}
where $\sigma_{excess}$ is the sum of any initial synchronization issues between the clocks, as well as any environmentally-induced and frequency drift-related timing offsets. In the recent state-of-the-art clock synchronization experiment, a ${\sigma_{excess}\simeq0}$ was achieved using classical frequency combs and sophisticated signal processing techniques (discussed further in \cref{sec:technical-history}) when ${0\leq\tau\leq1}$~s\cite{Caldwell2023}. Within this region, ${\sigma_{\Delta t}\equiv\text{TDEV}}$ and white PM noise dominated the system yielding a ${\text{TDEV}\propto1/\sqrt{\tau}}$ --- also referred to in the literature as the ``quantum-limited white noise floor''\cite{Caldwell2023,Caldwell2024}. TDEV and FFI measurements are at the SQL when white PM is the only noise source in the system. %Evidently, previous studies have shown that classical frequency combs can achieve the SQL over real free-space links\cite{Caldwell2023,Caldwell2024}. 

%State-of-the-art clock synchronization protocols can achieve $\sigma_{excess}\rightarrow0$ (see for example\cite{Caldwell2023}), and thus the focus of the next-generation of clock synchronization should be on minimizing TDEV.

%The TDEV provides a time-domain metric with units in time (often seconds) and represents the residual uncertainty after clock synchronization. The lower the final TDEV is, the lower the residual uncertainty, and the higher the overall precision of a given synchronization technique.
%When comparing different optical clock architectures and clock synchronization techniques we will use the FFI and specify the required $\tau$ as both quantities provide that are valuable for future implementations on satellites. 
In our perspective, future satellite-based clock synchronization techniques should aim to achieve an $\text{FFI}^{(2)}$ of ${<10^{-18}}$ within ${\tau\leq100}$~s for satellite-based clock synchronization. At this stability, a TDEV in the sub-femtosecond regime would be achievable using classical and quantum frequency combs (with femtosecond pulse duration) operating at the white noise floor. An integration time of less than $100$~s would also ensure compatibility with the typical viewing time-window of a low-Earth-orbit (LEO) satellite\cite{Zhang2019}.
%\textcolor{blue}{Technically, it represents the blah blah ???, and its relation to timing synchronization is found through ???. Here, will use the terms "FFI" and "uncertainty" interchangeably. [???suggest we just use FFI and FFIs as required?]}

\subsection{Optical clocks on ground and in space}
Current optical clock technologies can achieve an FFI of order $10^{-18}$ in ground-based laboratory settings, surpassing the performance of cesium-based atomic clocks. This level of stability has been achieved in experiments that are focused largely on lowering the FFI by overcoming technical issues such as second-order Doppler shifts\cite{Brewer2019}, thermal radiation\cite{Huntemann2016} and more\cite{Huntemann2016,Jian2023}. Some examples of architectures include the strontium ion optical lattice (FFI of $10^{-18}$ at $\tau=10^4$~s\cite{Bothwell2019} and $\text{FFI}^{(1)}$ of $4\times10^{-19}$ at $\tau=5\times10^{3}$~s\cite{Nicholson2015}), and the ytterbium ion optical lattice ($\text{FFI}^{(0)}$ of $4\times10^{-19}$ at $\tau=10^5$~s\cite{McGrew2018}). As the field of optical clocks matures, the focus will shift to achieving a shorter duration of $\tau$ for a given level of FFI, in addition to lowering the FFI. Furthermore, future optical clocks will be placed in orbit around the Earth to escape the noise effects due to Earth's gravitational fluctuations, which is a limiting factor for $\text{FFI}^{(2)}$ of $10^{-18}$ at $\tau\leq100$~s at present\cite{Riehle2017,Derevianko2022}. Satellite-based optical clocks will also aid spacecraft navigation by avoiding communication delays with Earth on corrections to trajectories. Optical clocks in space will also be used for the global dissemination of time and the re-definition of the second. However, current optical and atomic clocks have high SWaP demands --- satellite-based solutions will require miniaturization and radiation-hardening, and considerable research is already underway toward these directions\cite{Riehle2017}. 
Satellite-based optical and atomic clocks is a research direction that is receiving considerable attention by various space agencies which includes the National Aeronautics and Space Administration and their Deep Space Atomic Clock project\cite{Ely2018a,Burt2021,Seubert2022}, the European Space Agency and their Atomic Clocks Ensemble in Space\cite{Cacciapuoti2009,Meynadier2018}, the German aerospace center's project COMPASSO\cite{Schuldt2023a}, and the Chinese Space Laboratory's Tiangong-2 cold atomic clock\cite{Liu2018}. While these projects are using atomic clocks, the findings should translate to optical clock technologies. 
In the near-future, when a network of satellite-based optical clocks are deployed, we will require a clock synchronization scheme that is able to share time and frequency information with stability at an FFI better than the optical clocks themselves, and also within a reasonably short $\tau$ duration that is suitable for satellite links. 

\subsection{\label{sec:protocols-history}Clock synchronization schemes without frequency combs}
\begin{figure}
    \centering
    \includegraphics[width=\linewidth]{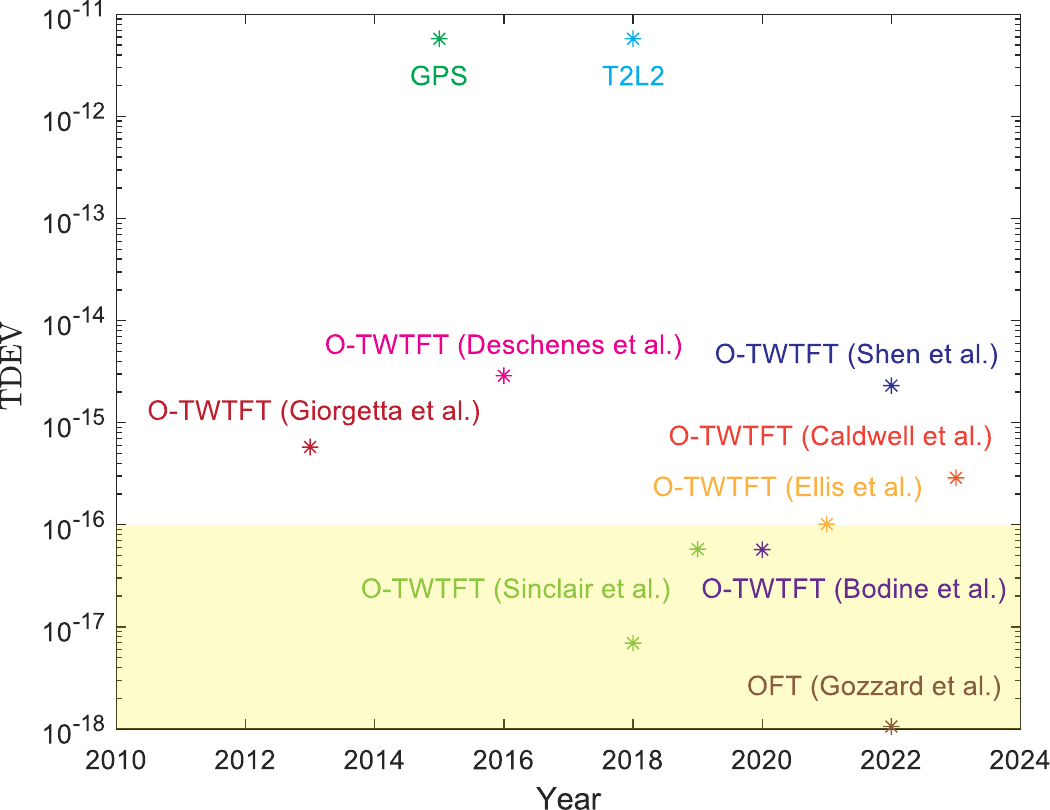}
    \caption{Historical comparison of different state-of-the-art free-space clock synchronization schemes. Note, the TDEVs plotted are the minimum values reported from the respective studies. In yellow we highlight the techniques that have achieved an $\text{FFI}^{(2)}$ of $\leq10^{-18}$ at $\tau\leq100$~s (i.e. a TDEV of $\leq10^{-16}$~s) which is our criteria for the performance required to synchronize satellite-based optical clocks.}% Acronyms used are as follows, %\textbf{O-TWTFT}: optical two-way time and frequency transfer, \textbf{GPS}: global positioning system, \textbf{T2L2}: time transfer by laser link, \textbf{OFT}: optical frequency transfer, and \textbf{TDEV}: timing deviation.}
    \label{fig:protocols-history}
\end{figure}
Current commercially available clock synchronization protocols, such as GPS common-view time transfer\cite{Kirchner1991,Kirchner1993} and GPS carrier phase\cite{Fujieda2014}, use a microwave carrier as the signal. However, although these techniques are suitable for synchronizing atomic clocks, they are unable to reach an FFI of $10^{-18}$ (and lower) which is required for a network of optical clocks\cite{Riehle2017}. For instance, the GPS carrier-phase protocol can achieve an $\text{FFI}^{(2)}$ of $10^{-16}$ at ${\tau=10^5}$~s\cite{Droste2015a}. Similarly, the carrier-phase two-way satellite time and frequency transfer schemes also suffers from the same threshold\cite{Hachisu2014,Fujieda2014,Bauch2015}. Whilst efforts are made to improve the FFI below this level\cite{Schafer2016}, the microwave-based carrier is a significant limiting factor. An opportunity exists for lowering the FFI by increasing the carrier frequency from microwave to optical --- this can be understood as an increase to $\nu_0$ in \cref{eq:yt,eq:xt} which would lower $y(t)$ and $x(t)$, in turn the FFI. In practice, however, optical signals are more sensitive to loss and noise in free-space than microwave signals which complicates the matter. Fortunately, recent experiments with optical signals are showing progress toward the required FFI levels. These experiments also reveal practical strategies for overcoming loss and noise issues that may be translated to satellite-based implementations in the near-future. %\textcolor{blue}{???why on the "protocol - are we mixing up protocols with hardware here?"} 

Time transfer by laser link (T2L2) is one such optical protocol that is based on the principles of a laser-ranging technique that collect TOF measurements\cite{SAMAIN2008}. In T2L2, several laser pulses are transmitted between a satellite and one or more ground stations, and detected using photodetection devices and time-tagging units. These laser pulses are short-duration and asynchronously transmitted from the ground stations to the satellite, with a fraction of them returned back to the ground stations. The ground stations record the start and return time of each pulse, while the satellite records the arrival time in the temporal reference frame of the on-board clock. A microwave carrier is then used to share the start, return and arrival times and calculate TDEV. Unfortunately, over real free-space links, the T2L2 protocol has under-performed thus far. 

In a recent field experiment, T2L2 was tested between the Jason~2 satellite and a single ground station\cite{Samain2015,Rovera2016}. In this particular test, the FFI achieved was of order $10^{-13}$ at $\tau=100$~s\cite{Samain2018}. Although, the FFI can be reduced to an $\text{FFI}^{(0)}$ of $10^{-17}$ at $\tau=1$~day\cite{Riehle2017}, this would be unsuitable for links LEO satellites. The main issue with T2L2 is that photodetection devices have a relatively slow response rate which limits the maximum sampling rate and in turn precision of the final TDEV measurements\cite{Shen2022}, despite the fact that in theory T2L2 should reach an $\text{FFI}^{(0)}$ of $10^{-17}$ at $\tau=300$~s\cite{Djerroud2010a} and an $\text{FFI}^{(2)}$ of $10^{-17}$ at $\tau=1$~s\cite{Robert2016}. %However, in practice, a detection scheme with faster response times and read-out would be needed. %Further improvements to T2L2 to overcome these practical challenges remain as open directions for research.

An alternative optical-based method is coherent optical frequency transfer (OFT). OFT uses an optical continuous-wave (CW) laser, instead of laser pulses (as in T2L2 and frequency comb-based techniques), that is phase-locked to a clock. The laser is transmitted to a remote site and interrogated with another CW laser that is locked to a different clock. By extracting the phase difference between the two CW laser signals, the TDEV between the clocks can be determined. The OFT protocol uses the phase information in the signal and, as a result, is highly sensitive to phase noise issues induced, for instance, by atmospheric turbulence. OFT is also sensitive to amplitude noise that reduce the signal-to-noise ratio and cause signal drop-outs. Amplitude noise can also occur due to turbulence and are a result of beam wandering, scintillation and deep fading in a channel\cite{Gozzard2018}. Without any noise-compensation strategies, phase and amplitude noise significantly limit the maximum $\tau$ achievable in any free-space clock synchronization technique, and in turn limit the FFI and TDEV. However, unlike pulse-based methods, OFT requires significant additional hardware for phase and amplitude noise suppression for successful operation. 

Recently, an OFT experiment conducted by Gozzard et al.\cite{Gozzard2022} used a $1532$~nm narrow-band ($<100$ Hz bandwidth) laser transmitted over a ${2.4~\text{km}}$ free-space terrestrial link. In this experiment, an $\text{FFI}^{(2)}$ of ${6.1\times10^{-21}}$ at ${\tau=300}$~s\cite{Gozzard2022} was achieved. This level of FFI exceeds the requirement for synchronizing optical clocks and satisfies our criteria for satellite-based clock synchronization (as shown in \cref{fig:protocols-history}). However, it is important to note that the performance strongly relies on the use of ultra-precise phase stabilization and amplitude noise suppression systems. Particularly, an imbalanced Michelson interferometer, an acousto-optic modulator, a bidirectional erbium-doped fiber amplifier and an adaptive optics unit were among some of the additional hardware components that were required to reduce the $\text{FFI}^{(2)}$ of ${10^{-15}}$ at ${\tau=10^4}$~s. In contrast, pulsed laser-based techniques have a large ``ambiguous range''\cite{Shen2021,Shen2022} which allow it to be more robust against atmospheric turbulence and signal drop-outs during deep fades\cite{Matthews2021}. For these reasons, the clock synchronization field at large has shifted focus to pulse-based techniques. 

A final point of interest is that both the T2L2 and the OFT present two different optical-based methods for clock synchronization. From a fundamental physics perspective, T2L2 follows the time-of-flight (TOF) method and relies on the precise comparison of the departure and arrival times of different sets of pulses to estimate the TDEV between two different clocks. Whereas OFT operates under the phase method and focuses instead on the relative phase difference between a local and remote narrow-band optical signal. Both methods, in theory, should be able to yield the same level of FFI (and in turn TDEV) for a given $\tau$ (discussed further in \cref{sec:quantum-fundamental}), but, in practice, require different hardware setups for operation over free-space links. As will be discussed in \cref{sec:classical,sec:quantum-combs}, classical and quantum frequency combs provide an alternative pathway for TOF and phase-based clock synchronization, and recent experiments are indicating that these approaches may also demand less hardware in practice. Less complex hardware should translate to lower SWaP demands which are highly desirable in space.
%An alternative scheme that boasts both low FFI and SWaP is classical frequency comb-based clock synchronization using a time-programmable frequency comb (TPFC). In a recent field experiment, a TPFC was successfully transferred over turbulent free-space $300$ km link achieving an FFI of $3.1\times10^{-19}$ after $\tau>10^3$~s and a TDEV at the SQL\cite{Caldwell2023}. Most importantly, this level of performance was achieved without the need of adaptive optics and other complex hardware components. The TPFC is therefore currently the most promising solution for satellite-based clock synchronization of an optical clock network \textcolor{blue}{[??? sentence makes no sense to me]}. 

A visual summary of the performance of various state-of-the-art clock synchronization techniques including a classical frequency comb-based technique known as optical two-way time and frequency transfer (O-TWTFT), discussed in \cref{sec:technical-history}, is presented in \cref{fig:protocols-history}, with our satellite-based clock synchronization criteria highlighted in yellow. %Here, we plot the TDEV, instead of the FFI, using \cref{eq:tdev} since the TDEV penalizes techniques that require longer $\tau$ for a given level of FFI. 

%\textcolor{blue}{
%[???not sure of of three things - the placement of this last paragraph  - and the use of tense changes in this section - the title of this section which i changed ,but not "history" in here]}
%\textcolor{red}{Removed last paragraph.}

\subsection{\label{sec:one-way-vs-two-way}One-way vs. two-way time transfer}
Another point of distinction between different clock synchronization techniques over free-space are the one-way and two-way transfer approaches\cite{Kirchner1991,Levine2008}. In the one-way approach, there exists one transmitter and one or more receivers. The transmitter could be a ground station or a satellite that generates a signal with timing and or frequency information about the transmitter clock. The receiver/s would capture the signal, decode and compare against the same information from a local clock. This approach is technologically simple and could cater for a large network, but there exist some implementation challenges. A challenge for the one-way method is that the location of the transmitter and receiver/s need to be known as precisely as possible before the exchange of signals takes place. The path distance between the transmitter and the receiver/s are needed to remove the path-based signal delay in the final calculation of TDEV. When the transmitter or receiver/s are not stationary, e.g. satellites, changes in path delay will contribute phase noise to the final TDEV estimate. Another key limiting factor for one-way is that random path-length variations caused, for instance, by atmospheric turbulence would also contribute phase and amplitude noise, as discussed in \cref{sec:protocols-history}. Finally, variations in the refractive index across the channel will effect the propagating speed of the signals also impact path-delay. For instance, the troposphere near the Earth's surface increases the effective path delay by $1$~ns per km\cite{Levine2008} which would need to be accounted for in ground-satellite links. One-way links are therefore only possible between stationary parties where the channel is well characterized \textit{a priori}. %GPS is an example of a one-way method where the path delay between satellite/s and ground station/s are estimated during triangulation\cite{Levine2008}.

Two-way transfer methods do not dependent on path knowledge in the same way\cite{Hanson1989}. Here, usually an intermediary satellite is used between two ground stations. The ground stations transmit signals to the satellite, and the satellite forward the signal from each ground station to the other. In theory, the position of any given satellite or ground station/s are no longer needed because all signals will travel from a ground station to the satellite and back, making path-based delays common across all signals. However, path-based noise will cancel only when the channel between the satellite and ground stations are reciprocal --- i.e., the effect of the channel in one direction can be canceled by propagation in the opposite direction. However, ground-to-satellite and satellite-to-ground channels are asymmetric since the former experiences the majority of atmospheric turbulence near the transmitter, while the latter experiences turbulence at the receiver. Despite this, recent modeling\cite{Robert2016,Belmonte2017} and experiments\cite{Swann2019} have found that a classical frequency comb-based technique known as optical two-way time and frequency transfer (O-TWTFT) would experience a negligible amount of noise over practical ground-satellite links due the short-duration pulsed nature of the frequency comb signal that is used. 

Both the one-way and two-way methods can be setup in various configurations which include exclusively between satellites without ground station/s. In this case, channel reciprocity may be satisfied due to there being negligible atmospheric turbulence in, for instance, LEO. Further, the simpler one-way method may also suffice in LEO links.

\subsection{\label{sec:technical-progress}Classical vs. quantum}
\begin{figure}
    \centering
    \includegraphics[width=\linewidth]{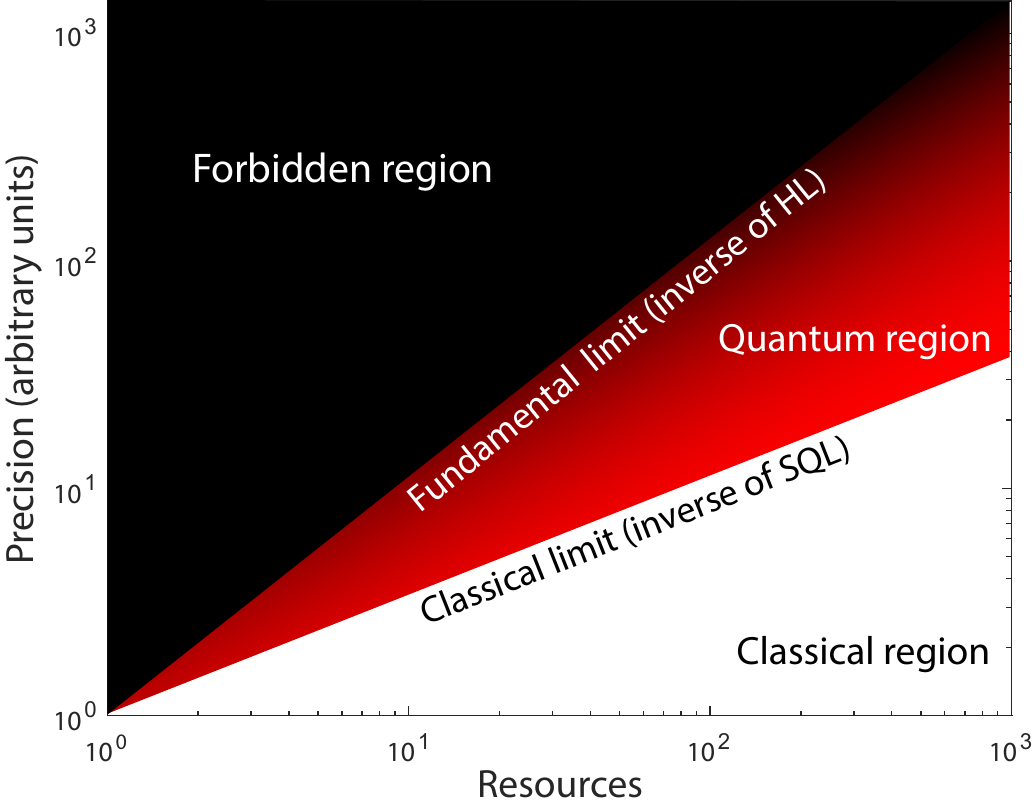}%,trim = 3.8cm 9.35cm 4.1cm 9.9cm, clip]{figure2.pdf}
    \caption{The precision vs. the number of resources (e.g. photons) trade-off for measurement strategies that use classical and quantum signals. All clock synchronization techniques that use the classical properties of light would operate in the classical region with a best-case performance at the SQL. On the other-hand, quantum frequency combs can exploit non-classical properties such as quadrature-squeezing and quadrature-entanglement to operate within the quantum region. We define performance anywhere in the quantum region as ``quantum advantage''.}% Acronyms used are as follows, \textbf{HL}: Heisenberg Limit and \textbf{SQL}: Standard Quantum Limit.}
    \label{fig:figure2}
\end{figure}
T2L2, OFT and O-TWTFT all use properties of light that pertain to classical physics. The best-case FFI (and TDEV) achievable by any of these ``classical'' techniques will be the SQL\cite{Giovannetti2001}. The SQL is technically a scaling in the standard deviation of a measurement given by $\sigma\propto1/\sqrt{n}$, where $\sigma$ is the standard deviation and $n$ is the number of resources used. In the context of clock synchronization, the FFI (and TDEV) when classical techniques are employed scale as $1/\sqrt{n}$ with $n$ being the number of signal photons captured during $\tau$.% Note that the SQL is achieved only after all noise sources (such as the amplitude and phase noise discussed in \cref{sec:protocols-history,sec:one-way-vs-two-way}) are entirely suppressed. The SQL thus represents the ultimate limit when using classical states of light.

Quantum physics provides a new scaling of the standard deviation by using ``non-classical'' correlations within the properties of light\cite{Giovannetti2002,Giovannetti2004,Giovannetti2011,Demkowicz-Dobrzanski2012,Napolitano2011a,Zwierz2012,Zhou2018,Napolitano2011}. Quantum states can be used to approach the HL, which is a scaling in the standard deviation given by $\sigma\propto1/n$, a $\sqrt{n}$ improvement relative to the SQL. There are two ways to interpret the HL: (1) for a given level of $\sigma$, a quantum state may use less $n$ than a classical, or (2) for a given amount of $n$, a quantum state may achieve a lower $\sigma$ than a classical. We define the inverse of the standard deviation as the precision. %, thus the uncertainty-to-resource trade-off can also be interpreted as a precision-to-resource trade-off. Note, that $n$ and $\tau$ are proportional such that lowering $n$ would decrease $\tau$.

To see the distinction between the SQL and the HL, we have presented a visualization of the precision vs. resources in \cref{fig:figure2}.
In resource-constrained settings such as on-board satellites, the precision-to-resource trade-off is an important consideration. In this work, we define a performance anywhere within the quantum region in \cref{fig:figure2} as ``quantum advantage'', because higher precision is achieved for a given number of resources relative to classical.%In this sense, using a quantum state may help to maximize the overall system efficiency. However, some caveats do exist in practice --- for instance, the opportunity for an advantage to using quantum states (i.e. a ``quantum advantage'') grows with the number of resources and so would be almost negligible when a small number of resources are available. Additionally, achieving a quantum advantage in free-space encompasses a distinct set of technical challenges which are discussed further in \cref{sec:quantum-combs}. All in all, the opportunity that quantum presents for clock synchronization could be substantial and one that, in our perspective, deserves attention. 

%However, in practice, the HL itself is challenging to obtain as the performance of quantum states of light are highly sensitive to photonic loss and noise\cite{Demkowicz-Dobrzanski2012,Gosalia2022,Gosalia2023}.  Although, as the overall number of resources increases, the difference between the SQL and the HL also grows --- distinguished by the quantum region in \cref{fig:figure2}. Since operation anywhere in the quantum region has the potential to provide a precision-to-resource advantage --- performance at the HL is not the only option. For instance, a setup using quantum squeezing would require infinite squeezing to reach the HL, which is impossible in practice. However, finite levels of squeezing, the state-of-the-art being $20$~dB\cite{Hosten2016}, would still overcome the SQL and provide a measurable precision-to-resource advantage with increasing $n$. The current focus of quantum-enhanced clock synchronization as a research direction is therefore in answering the question: can quantum states of light be efficiently generated and exchanged to retrieve a precision-to-resource advantage over free-space SWaP-constrained links? In \cref{sec:quantum-combs}, we further discuss this question and provide some insights based on recent work.

%[??? I am finding this discussion on resources and different limits somewhat cumbersome and/or confusing ]

\section{\label{sec:classical}Classical frequency combs}
In this section, we provide focus on clock synchronization using classical frequency combs, and discuss the generation and operating principles of a mode-locked laser (MLL) --- a popular choice for generating classical frequency combs. We also discuss some challenges towards noiseless operation as well as some findings from recent satellite-based experiments with MLLs. This section serves as an overview of the field in general and outlines progress toward SWaP optimizations.

\subsection{Generating and characterizing classical frequency combs}
\begin{figure}
    \centering
    \includegraphics[width=\linewidth]{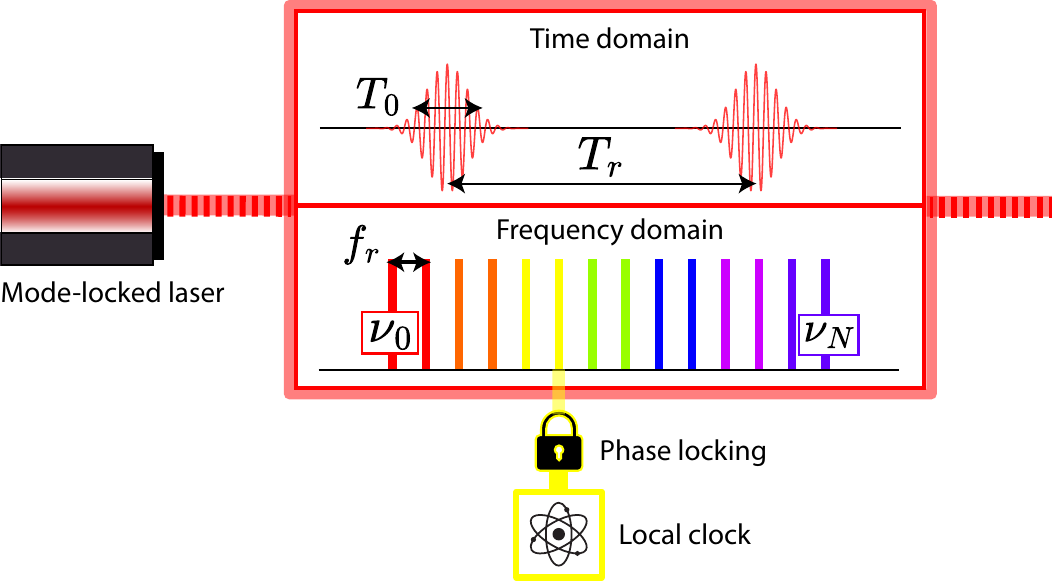}
    \caption{The frequency and time domain representations of a classical frequency comb based on an MLL. In the optical frequency domain, the comb is a collection of optical modes denoted $\nu_N$ with mode spacing $f_r$ (the repetition rate of the laser). In the time domain, the comb is a periodic train of optical pulses with pulse period $T_r=1/f_r$. For clock synchronization, one of the frequency comb modes will be first phase-locked to the transition of an optical clock. Due to the coherence shared between all frequency comb modes, all modes will in turn become phase-locked to the optical clock as well. Then, the frequency comb will be transmitted over free-space to a remote site. At the remote site, interferometry is used to compare the transmitted comb with a ``receiver'' comb (that is phase-locked to a receiver optical clock). Interferometry will reveal differences in the phase and/or TOF information between the two combs which is in turn linked to the TDEV between the transmitter's and receiver's clocks\cite{Giorgetta2013}.}% Acronyms and symbols used are as follows, \textbf{MLL}: mode-locked laser, \textbf{TOF}: time-of-flight, \textbf{TDEV}: timing deviation, $T_0$: pulse duration, $T_r$: pulse period, $f_r$: repetition rate, $\nu_N$: optical frequency at index $N$.}%\textbf{(a)} Frequency and time domain representations of a frequency comb. In the optical frequency domain, the comb is a collection of optical modes denoted $\nu_N$ with mode spacing $f_r$ (also referred to as the repetition rate of the laser). In the time domain, the comb is a periodic train of optical pulses with pulse period $T_r=1/f_r$. The optical modes have two degrees of freedom: $f_r$, and the carrier-envelope offset frequency $f_0$; $\nu_N=N f_r+f_0$, where $N\in\mathbb{Z}$ is the mode index. For stabilization, both of these degrees of freedom need to be measured and controlled. To measure $f_r$, typically an optical photodetector is used to detect the amplitude modulation of the optical pulse train resulting in an electronic pulse train of coherently related microwave Fourier harmonics at $f_n=nf_r$ where $n\in\mathbb{Z}$. \textbf{(b)} While $f_r$ is measured at microwave frequencies, $f_0$ needs to be measured at optical frequencies by the $2v_N-v_{2N}$ self-referencing method (also called $f-2f$ interferometry\cite{Cundiff2001}). In the yellow box, the effect of $f_0$ on the carrier-envelope phase, $\phi_{\text{CEO}}(t)$, is shown.}
    \label{fig:freq-comb}
\end{figure}
\begin{figure}
    \centering
    \includegraphics[width=\linewidth]{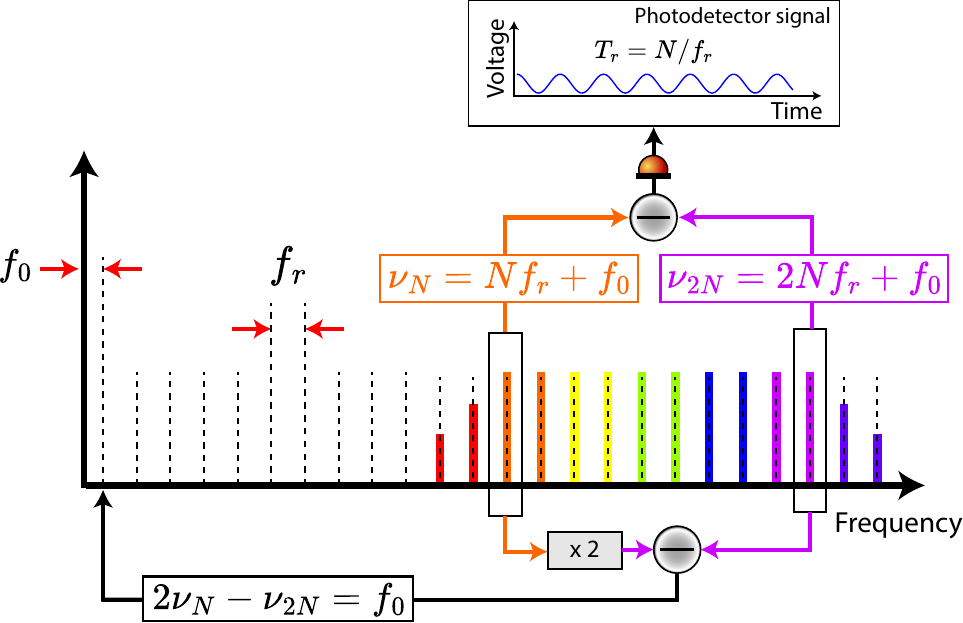}
    \caption{A visualization of the MLL-based frequency comb stabilization process. At the bottom, the self-referencing technique for determining the offset frequency, $f_0$, is shown. At the top, we show the amplitude modulation technique which consists of an optical photodetector that is used to extract an electronic signal whose period, $T_r$, is related to the frequency comb mode spacing $f_r$. Real MLL outputs exhibit an approximately Gaussian shape in the frequency domain centered at an optical frequency with at least a bandwidth that spans atleast one octave (for self-referencing)\cite{Giorgetta2013}.}% Acronyms and symbols used are as follows, \textbf{MLL}: mode-locked laser, $T_r$: pulse period, $f_r$: repetition rate, $f_0$: offset frequency, $\nu_N$: optical frequency at index $N$, and $N$: frequency mode index.}
    \label{fig:freq-comb-stabilization}
\end{figure}
In the literature, the term (classical) ``frequency combs'' is often synonymous in usage with ``MLL''\cite{Diddams2010a}. This is because MLLs were the original source of classical frequency combs in the early 2000s when they were used to perform read-outs of an optical clock by coherently and precisely converting the clock's optical cycles into measurable microwave signals\cite{Haus2000,Cundiff2001b,Cundiff2003a}. Since then, applications for MLLs have widened and techniques for generating and stabilizing MLLs has also matured considerably\cite{Jin2016,Ma2019,Fortier2019,Diddams2020}. MLL-based combs remain a popular choice for comb generation, with the focus recently shifting toward balancing the trade-off between producing a wide spectral bandwidth, high frequency resolution and a low SWaP footprint, on the road to widespread commercialization\cite{Chang2022}. Although there are other methods for generating classical frequency combs that are based on nonlinear processes such as Kerr combs, electro-optic combs, and quadratic combs\cite{Ricciardi2020,Kovach2020,Wang2020,Hu2021,Chang2022}, we focus mainly on MLL-based classical frequency combs. 

An MLL-based frequency comb start as a single-frequency continuous-wave laser that is passed through a resonant cavity and stabilized via active or passive mode-locking. The optical-frequency modes of this comb are mode-locked during the stabilization process which ensures that they are equidistant in the frequency domain, phase-coherent in the time domain, and share a common phase evolution that is deterministic\cite{Fortier2019}. %The output pulses of an MLL-based comb are visualized in \cref{fig:freq-comb}. 
Mode-locking is a resonant phenomenon that has been developed over many decades\cite{Haus2000,Smith1970,Smith1974}. In principle, a single-frequency laser is converted into a train of ultra-short duration pulses by repeatedly passing the laser through a resonator in a cavity. Each cavity round-trip produces pulses that shorten in the time domain, with successive passes, and simultaneously widen in the frequency domain. The round-trips continue until the cavity characteristics prevent further time-shortening and spectrum-expansion and, at this point, the pulses escape the cavity\cite{Haus2000}. MLLs are typically ultra-short in duration ($T_0\simeq10^{-14}$~s, where $T_0$ is the pulse duration as shown in \cref{fig:freq-comb}), have a wide spectral bandwidth (order 1000~nm for octave-spanning combs), and relatively high peak output power (order 10~mW). These three characteristics are highly desirable for high-precision sensing applications including clock synchronization\cite{Hall2006,Lamine2008}. 

In principle, MLLs are simple to operate as there are only two parameters in the frequency domain that can be tuned. These parameters are: the laser repetition rate ($f_r$) and the carrier-envelope offset frequency ($f_0$). These two parameters can be considered as the two degrees of freedom of a frequency mode, denoted by $\nu_N$, since
\begin{gather}
    \nu_N=Nf_r+f_0,
    \label{eq:comb-equation}
\end{gather}
which is referred to as the ``frequency comb equation'', and $N\in\mathbb{Z^+}$ is the mode index\cite{Diddams2010a,Fortier2019}. We can equivalently describe an MLL in the time-domain, but in this case we have three parameters: the pulse duration ($T_0$), the pulse-to-pulse period ($T_r=1/f_r$), and the pulse-to-pulse phase slip between the carrier and envelope ($\Delta\phi_{ceo}=2\pi f_0/f_r$). In \cref{fig:freq-comb,fig:freq-comb-stabilization}, we have provided visuals describing these parameters to aid. Ultimately, the sustained operation of an MLL requires tight monitoring and control of the aforementioned parameters during the generation stage --- the tighter the control, the more stable the generated pulses are, and the more precise the FFI (and TDEV) will be over free-space. 

\subsection{\label{sec:classical-noise}Stabilizing MLLs and suppressing noise}
The parameters of an MLL are typically measured and controlled using feedback loops\cite{Fortier2019}. Firstly, to measure $f_r$, an optical photodetector is used to detect the amplitude modulation of a group of pulses (referred to as a pulse train) which in turn produces an electronic train of microwave Fourier harmonics at $f_n=nf_r$ where $n\in\mathbb{Z}^+$. One example where only two modes were interrogated is provided in \cref{fig:freq-comb-stabilization}, although in practice the entire spectrum of modes could be analyzed\cite{Giorgetta2013}.
Secondly, $f_0$ is measured at optical frequencies by the self-referencing method (also called $f-2f$ interferometry\cite{Cundiff2001}). In sensing applications, it is essential that $f_r$, $f_0$, $T_r$, $\Delta\phi_{ceo}$ and $\tau$ are stable during operation. Specifically in the clock synchronization context, it is highly desirable that $10$~MHz~$\leq~f_r~\leq~1$~GHz, as this range allows for a commercial-of-the-shelf photodetection units for use in measurement schemes such as linear optical sampling (LOS)\cite{Dorrer2003,Fortier2019}. 

Unfortunately, due to various sources of noise that can arise during the generation stage, real-time measurements of the parameters (i.e. $f_r$ and $f_0$ mainly) are required for optimum performance. A negative feedback loop based on tuning the cavity length is one common approach for noise suppression\cite{Kim2016,Tian2021}. For instance, in a fully-stabilized MLL (where the parameters are fixed), $T_r$ is precisely the round-trip time for a single pulse through the laser cavity. However, changes in the cavity length even down to an optical wavelength in length (i.e. nanometer-scale) will impact the $T_r$ of the output MLL, causing a phenomenon termed timing jitter\cite{Kim2016,Tian2020}. Cavity length changes will impact the temporal spacing between successive pulses as they leave the cavity.
Timing jitter is unfortunately a common issue and can arise due to many factors, here we list some common sources and strategies. Sources of timing jitter include intra-cavity amplified spontaneous emission\cite{Herda2005,Liao2020}, cavity dispersion\cite{Mulet2006}, intensity fluctuations\cite{Wang2019}, and slow saturable absorber recovery time\cite{Bao2004} (in passive mode-locked combs). 
Several techniques for controlling timing jitter also exists and these include the phase detector method\cite{Mao2020}, balanced optical cross-correlation\cite{Song2011,Li2012}, optical heterodyne\cite{Hou2015}, delayed optical heterodyne\cite{Tian2020}, intrinsic noise optimization\cite{Wang2019} and phase locking\cite{Briles2010}. 
Recently, the balanced optical cross-correlation method is gaining popularity due to its superior performance\cite{Song2011}. 

Over the last few decades, models for timing jitter-related noise have improved considerably\cite{VonderLinde1986,Haberl1991,Haus1993,Namiki1997,Paschotta2004}. These models have helped to focus future noise suppression strategies by revealing the main contributing factors of timing jitter. Namely, it was found that shortening the $T_0$ (the pulse duration) and eliminating cavity dispersion are the best strategies. Shortening $T_0$ helps to reduce the impact of amplified spontaneous emissions, while eliminating cavity dispersion helps to minimize the intensity- and frequency-related fluctuations in the output MLL\cite{Song2011}. Using these insights, Song et al.\cite{Song2011} were able to produce an MLL with an ultra-low timing jitter of $175$~attoseconds on a ytterbium fiber-based frequency comb. Their noise-suppression technique was based on balanced optical cross-correlation\cite{Song2011} however, due to limitations in their setup, the output MLL had a maximum $f_r\leq80$~MHz since the timing jitter was considerably worse at higher $f_r$. More recently, Ma et al.\cite{Ma2018}, was able to extend this performance on the same ytterbium-fiber comb to an $f_r\leq750$~MHz using an electro-optic modulator and a piezo-electric transducer units, and have also inspired other works\cite{Deng2021}. 
%Although timing jitter is potentially a significant issue for clock synchronization, this experimental outcome shows that timing jitter is well understood and controllable to the order 100 attosecond level. We expect that the lessons learned will also carry to other types of classical combs, and quantum frequency combs in the near future. Especially in clock synchronization, timing offset measures rely heavily on a stable $T_r$. This is because in interferometric measurement schemes like LOS, small differences in $T_r$ between two or more combs is used to estimate the timing offsets between clocks. Deviations due to timing jitter noise would cause error in this timing offset measurement, and thus reduce the overall clock synchronization precision. 

Since the $T_0$ of a typical MLL is at the order of the duration of a few optical cycles, instabilities in $\Delta\phi_{ceo}$ and in turn the $f_0$ (referred to as phase noise) are common\cite{Tian2021}. Particularly, fluctuations in $\Delta\phi_{ceo}$ during operation would induce relative phase-shifts between pulses reducing the coherence between pulses. These fluctuations are a result of changes between the group and phase velocity of a pulse as it transits the cavity and can arise due to intra-cavity amplified spontaneous emission, cavity loss, pump noise, and fluctuations in the laser cavity length\cite{Herda2005,Liao2020,Deng2021,Tian2021}. Although self-referencing is a useful method for measuring changes in $f_0$, practical implementations have some complexities that include the requirement of an octave-spanning bandwidth, perfect mode-matching of the $2\nu_N$ and $v_{2N}$ modes and strict temperature control \cite{Endo2018a}. %Impressive levels of $f_0$ stabilization have been achieved in the laboratory, with a state-of-the-art FFI $9.15\times10^{-19}$ at $\tau=1$~s\cite{Deng2021}.
Impressive levels of timing jitter and phase noise control have been achieved in recent laboratory experiments\cite{Ma2018}, which have shown that it is possible to operate an MLL with an $\text{FFI}^{(2)}$ of ${3\times10^{-19}}$ at ${\tau=10^3}$~s\cite{Deng2021}. %In the next section, we discuss recent field experiments where these highly-stabilized MLLs have been successfully transfered over free-space using the O-TWTFT protocol. 

\subsection{SWaP optimization with integrated photonics}
A satellite-based implementation of MLL will require significant miniaturization and space-hardening from their current form. A key challenge would be to ensure that there is no degradation in performance as a result --- toward this, the field of integrated photonic is paving a potential pathway. In a recent review by Chang et al.\cite{Chang2022}, integrated photonic is discussed as an emerging field of development for producing high-volume, low-cost, and SWaP-efficient photonic hardware based on common materials like silicon (which is widely available)\cite{Thomson2016}. 

Current technological directions are focused on replacing individual optical components with their equivalent integrated photonic-solutions one component at a time, and testing these devices on a working laboratory setup. For example, Carlson et al.\cite{Carlson2017} developed an integrated photonic circuit based on silicon nitride that uses multiple waveguides which are all excited by a single (externally generated) MLL pump. The waveguides produce super-continuum light (spectral-broadened) at wavelengths that corresponded to different optical clock standards. The super-continuum light overlaps with lasers locked to various optical clocks, and heterodyne measurements are conducted to estimate the TDEV between all clocks using the waveguide output. This particular integrated photonic device replaces a bulky spectrum widening apparatus with a low-SWaP chip-based solution. The performance metric achieved includes a residual FFI of $3.8\times10^{-15}$ at $\tau=2$~s\cite{Carlson2017}, which is progress in the right direction. Another study by Jankowski et al.\cite{Jankowski2020} consists of both a photonic chip-based second harmonic generation stage and a chip-based spectral broadening stage were demonstrated. These chip-based devices showed power efficiencies as they consumed order femtojoules of energy  ---  a reduction of 3 orders of magnitudes from previous laboratory-based setups. As the technology of integrated photonic continues to mature, similar SWaP optimizations could be expected for all the various components used in clock synchronization\cite{Gaeta2019,Wang2020,Chang2022}.

%\textit{\textcolor{blue}{To add: recent developments in generating classical frequency combs on integrated photonic: different techniques, compare performance, and suggest potential directions of future research.}}
%In terms of chip-level generation of frequency combs, the two main ...

\subsection{Experiments in space}
\begin{table*}[t]
    \centering
    \caption{A comparison of recent satellite-based experiments and proposals using MLL-based frequency combs. The aim of this table is to suggest similarities between certain attributes, and also point out some parameters that are often missing in the literature, but are crucial for an assessment of space-readiness. Note the symbols below include $P_{in}$: the input pump power, and $P_{out}$: the output seed power.}% Acronyms and symbols used are as follows, \textbf{Er}: Erbium, \textbf{MLL}: mode-locked laser, $f_r$: repetition frequency, $T_0$: pulse duration, $P_{in}$: input pump power and $P_{out}$: output seed power.}
    \label{tab:mll-space-experiments}
    \begin{tabular}{c c c c c c c c c c}
        \Xhline{3\arrayrulewidth}
         Project (Year) & Technology & Pump (nm) & Seed (nm) & $f_r$ (MHz) & $T_0$ (fs) & $P_{in}$ (mW) & $P_{out}$ (mW) & Volume (L) & Weight (kg)\\
         \Xhline{2\arrayrulewidth}
         STSAT-2C\cite{Lee2014} (2014) & Er:fiber & $980$ & $1590$ & $25$ & $350$ & $600$ & $14$ & $3.3$ & $2.5$\\
         FOKUS I\cite{Lezius2016} (2016) & Er:fiber & $780$ & $1560$ & $100$ & - & 10 & - & $15$ & $20$ \\
          FOKUS II\cite{Lezius2019} (2021) & Er:fiber & $980$ & $1560$ & $100$ & $45$ & - & $180$ & $7$ & $10$\\
          \textit{proposed}\cite{Takeuchi2021} (2021) & Er:fiber & $976$ & $1560$ & $48.7$ & - & - & $0.8$ & - & - \\
          \Xhline{3\arrayrulewidth}
    \end{tabular}
\end{table*}

To our knowledge, there have recently been three space-based experiments that have demonstrated the operation of MLL-based frequency combs in space. Although these experiments used a single spacecraft (satellite or sounding rocket), the findings shed some insights on the key challenges and opportunities that lay ahead as the field transitions to large-scale network of free-space MLL links (as shown in \cref{fig:figure1}).
In a recent mission in collaboration between the Korean Advanced Institute of Science and Technology and the South Korean Satellite Technology Research Center, an Erbium fiber (Er:fiber) MLL was tested onboard the satellite STSAT-2C\cite{Lee2014}. This particular MLL emitted pulses with $T_0=350$~femtosecond, an $f_r=25$~MHz, and was centered at a $1590$~nm wavelength.
During the mission, the MLL successfully endured the high-accelerations during launch\cite{Baumann2009,Sinclair2014} as well as high-energy space-radiation\cite{Fox2013} with limited changes to the MLL parameters.
Lee et al.\cite{Lee2014} showed that the MLL could sustain continuous operation over one year and, for instance, achieve an $f_r$ FFI of $10^{-12}$ at $\tau=10$~s\cite{Lee2012} throughout the year.
This performance was despite an $8.6\%$ reduction in the average MLL output power level which is attributed to radiation-induced attenuation.
For a complete description of the setup please see\cite{Lee2014}; we briefly highlight that the comb stabilization setup included a saturable absorber (for passive mode-locking that was directly inserted to avoid using bulk optics) and a ring-type piezoelectric actuator system (to stabilize $f_r$). 
%In this setup, the timing jitter noise was actively controlled by varying the cavity length using a ring-type piezoelectric transducer.
%Notably, these performance metrics were met despite external noise contributions from on-board vibrations, temperature variations and radiation, although these metrics are significantly worse than the state-of-the-art laboratory-experiments\cite{Benedick2012a}, as discussed in \cref{sec:classical-noise}. 
As future recommendations, some hardware suggestions were made including the use of an active temperature control unit for greater noise reduction, and thicker aluminum shielding for radiation proofing. Although these suggestions would increase the overall SWaP, the use of integrated photonic could provide alternative options.

In the FOKUS I mission a substantially shorter experiment was conducted with the sounding rockets that lasted $360$~s in total\cite{Lezius2016}. An Er:fiber MLL was again used, but this time to continuously compare the TDEV between an optical and microwave clock. Specifically, MLL was phase-locked to the $384$~THz optical transitions of a rubidium optical clock, and then compared with the transitions of a cesium atomic clock that oscillates at $10$~MHz via the $f_r$ and $f_0$ frequency comb parameters. The performance achieved during the operation was an FFI of $10^{-11}$ at $\tau=20$~s\cite{Lezius2016}, however the experimental duration was too short for any further meaningful conclusions. A successive mission, the FOKUS II, took the lessons learnt from FOKUS I and used a testing payload with lower SWaP than the FOKUS I\cite{Probster2021}. Further, a dual-comb setup was used with two MLLs to compared TDEV between an iodine-based optical clock at $281$~THz optical transition and a cesium atomic clock. The FOKUS II achieved an $\text{FFI}^{(1)}$ of $4\times10^{-12}$ at $\tau=100$~s\cite{Probster2021}. Although, since the FFI method used in the FOKUS I mission was not made clear, we cannot easily compare the performance between both missions. Nevertheless, as an early demonstration, both the FOKUS I and II missions show that MLLs can be used to autonomously and continuously compare between different clock technologies in space, while sustaining reasonably low levels of FFI.  

%In principle, the overall performance of a satellite-based MLL frequency comb will depend on the long-term stability of the mode-locked properties and the sustained control of typical noise factors. A satellite-based deployment's primary challenge is achieving high stability and low-noise performance in a compact, space-hardened, and energy-efficient package.
A final space-based experiment that is currently under development is by Takeuchi et al.\cite{Takeuchi2021}. The setup proposed in this experiment also uses an Er:fiber MLL in a ``figure-8'' configuration with a nonlinear polarization rotation device used for stabilization. Early ground-based tests conducted continuously over several days are showing stable operation with FFI of $10^{-14}$ at $\tau=2000$~s\cite{Takeuchi2021}.
A summary of all satellite-based experiments are provided in \cref{tab:mll-space-experiments}. Due to the limited standardization in the field currently, some entries are missing --- as the field matures in coming years, we encourage reporting of these metrics for greater ease of assessment of space-readiness.

%In \cref{tab:mll-space-experiments}, we compare the performance of recent satellite-based experiments and proposals using MLL frequency combs. Due to limited standardization in the field, some entries are missing from the literature. 

\subsection{\label{sec:technical-history}Optical two-way time and frequency transfer}
The most prominent MLL-based clock synchronization technique is O-TWTFT. First proposed by Giorgetta et al.\cite{Giorgetta2013}, O-TWTFT combines the lessons learned from fiber-based time-frequency transfer\cite{Predehl2012,Droste2013,Bercy2014a,Ning2014} and the microwave-based two-way time-frequency transfer\cite{Bauch2015}. In O-TWTFT, two sets of MLL pulses are locally produced and phase-locked to two different clocks. These MLL pulses are then exchanged over free-space in a two-way configuration and compared via an interferometric method such as LOS\cite{Dorrer2003}. Through interferometry, the TOF or phase differences between the two sets of pulses are compared, which corresponds to the degree to which the two clocks are out of sync. A series of experiments over progressively longer free-space terrestrial links have been conducted using O-TWTFT in recent times\cite{Giorgetta2013,Deschenes2016a,Sinclair2016,Sinclair2018a,Bodine2020,Shen2022}. The results from these experiments, and those from some previous fiber-based methods are summarized in \cref{tab:classical-comb-terrestrial} for ease of comparison.

\begin{table*}[t]
    \centering
    \caption{Recent field experiments using MLL-based classical frequency combs over optical fiber and terrestrial links and their reported performance metrics. The general trend across the recent free-space experiments has been toward longer link range, shorter integration time ($\tau$), and an $\text{FFI}^{(2)}$ below $10^{-18}$ (in turn, lower TDEV). Note, \textbf{OPD} is optical phase detection\cite{Bercy2014a}.}%  Acronyms used are as follows, \textbf{FFI}: fractional frequency instability, \textbf{TDEV}: timing deviation, \textbf{OFT}: optical frequency transfer, \textbf{OPD}: optical phase detection, and \textbf{O-TWTFT}: optical two-way time and frequency transfer.}
    \label{tab:classical-comb-terrestrial}
    \begin{tabular}{c c c c c c c c c}
        \Xhline{3\arrayrulewidth}
         Author & Year & Protocol & Link type & Nodes & Path (km) & $\text{FFI}^{(2)}$ & Integration time, $\tau$ (s)~~ & TDEV (s)\\
         \Xhline{2\arrayrulewidth}
         Predehl et al.\cite{Predehl2012} & 2012 & OFT & fiber & $11$ & $920$ & $10^{-18}$ & $10^3$ & $6\times10^{-16}$ \\
         Droste et al.\cite{Droste2013} & 2013 & OFT & fiber & 2 & $1840$ & $4\times10^{-19}$ & $10^2$ & $2\times10^{-17}$ \\
         Bercy et al.\cite{Bercy2014a} & 2014 & OFT & fiber & 2 & $100$ & $5\times10^{-21}$ & $10^3$ & $3\times10^{-18}$ \\
         %Ning et al.\cite{Ning2014} & 2014 & OPD & Fiber & 2 & $10$ & $8.8\times10^{-19}$ & $4\times10^4$ & $3.5\times10^{-15}$ \\
        Giorgetta et al.\cite{Giorgetta2013} & 2013 & O-TWTFT & free-space & 2 & $2$ & $10^{-18}$ & $10^3$ & $6\times10^{-16}$ \\
         Deschenes et al.\cite{Deschenes2016a} & 2016 & O-TWTFT & free-space & 2 & $4$ & $5\times10^{-19}$& $10^4$ & $3\times10^{-15}$\\
         Sinclair et al.\cite{Sinclair2018a} & 2018 & O-TWTFT & free-space & 2 & $4$ & $10^{-17}$ & $1$ & $7\times10^{-18}$\\
         Sinclair et al.\cite{Sinclair2019} & 2019 & O-TWTFT & free-space & 2 & $4$ & $10^{-18}$ & $10^2$ & $6\times10^{-17}$\\
         Bodine et al.\cite{Bodine2020a} & 2020 & O-TWTFT & free-space & 3 & $ 14$ & $10^{-18}$ & $10^2$ & $6\times10^{-17}$\\
         Ellis et al.\cite{Ellis2021a} & 2021 & O-TWTFT & free-space & 3 & $28$ & $10^{-18}$ & $2\times10^2$ & $1\times10^{-16}$\\
         Shen et al.\cite{Shen2022} & 2022 & O-TWTFT & free-space & $2$ & $113$ & $4\times10^{-19}$ & $10^4$ & $2\times10^{-15}$\\
         Caldwell et al.\cite{Caldwell2023} & 2023 & O-TWTFT & free-space & 2 & $300$ & $5\times10^{-18}$ & $1\times10^3$ & $3\times10^{-16}$\\
         \Xhline{3\arrayrulewidth}
    \end{tabular}
\end{table*}

From \cref{tab:classical-comb-terrestrial}, it is clear that there have been a large number of O-TWTFT experiments in recent years. However, the experiments conducted from 2013--2021 had limited free-space range (i.e. less than $30$~km). A key obstacle in these experiments has been the LOS method and, in particular, the relatively high signal strength (order nanowatts) required for optimal photodetection which is a key part of conventional LOS. 
Some technical strategies were proposed to overcome this issue which have focused on meeting the power demand by increasing the transmit power, using a larger receiver telescope aperture, or using an adaptive optics setup\cite{Shen2022}. However, these methods all add to the SWaP costs of the overall setup and would thus have limited prospects in space.

Fortunately, a SWaP-friendly direction has been recently developed that focuses instead on improving the LOS method by using something called a time programmable frequency comb (TPFC)\cite{Caldwell2022,Caldwell2022b,Caldwell2023,Caldwell2024}. The TPFC is a highly-precise MLL that can be tuned in time and phase in a coherent manner with sub-$10$~attosecond accuracy\cite{Caldwell2022,Caldwell2022b}. In traditional LOS\cite{Dorrer2003,Giorgetta2013,Deschenes2016a,Sinclair2014,Bodine2020,Ellis2021a}, a signal MLL is sampled using another ``local'' MLL using heterodyne detection. Instead of a local MLL, a TPFC can be used to provide finer-tuning capabilities during sampling. Using a TPFC greatly improves the accuracy of measurements, and has been shown to enable reductions in the required receiver power level from nanowatts to order $0.1$~picowatts\cite{Caldwell2023}.  

Over a $300$~km terrestrial link, the TPFC-based O-TWTFT was used to synchronize two distant clocks to a best-case TDEV of $500$~attoseconds, which was at the SQL for this particular setup\cite{Caldwell2023}. The achievable $\text{FFI}^{(2)}$ of ${3\times10^{-19}}$ at ${\tau=10^3}$~s\cite{Caldwell2023}. This performance is near the state-of-the-art considering that the free-space link experienced $102$~dB loss and the median received signal power level was only $0.15$~picowatts\cite{Caldwell2023}. Moreover, the TPFC was not pre-amplified at the transmitter, and this performance level was achieved without adaptive optics. Notably, Caldwell et al.\cite{Caldwell2023} also demonstrated a Kalman filter to continuously monitor the TDEV between the local and signal MLL during the absence of deep fades. %As mentioned already, the TPFC precision was also shown to approach the SQL in performance, for the first time, and is the ultimate limit for classical frequency combs. 
%A summary of recent field experiments using classical frequency combs over terrestrial links and their corresponding performance metrics are given in \cref{tab:classical-comb-terrestrial}. For comparison, state-of-the-art performance of optical clock synchronization over fiber links are also provided.
%O-TWTFT experiments have recently focused on terrestrial free-space links as proof-of-concept with the aim of future expansion to satellite-based links. Fractional frequency instability (modified Allan deviation) Balance received optical power, timing performance and update rate, extend link distance at constant launch power and telescope aperture trade-off in sensitivity, speed and resolution

In summary, there are some promising directions that have been laid for clock synchronization using classical frequency combs. Recent developments in the field have demonstrated that using optical signals, instead of microwave, allows for achieving a lower FFI with shorter $\tau$ which are essential for satellite-based optical clock synchronization. Further, using a pulsed signal rather than CW offers a more robust method against turbulent channels. A TPFC-based system also enables a longer link with limited additional complex hardware, and would be a SWaP-friendlier option. Further, the TPFC-method is able to reach the SQL in performance, and thus may be a good solution for satellite-based deployment in the future. In our perspective, integrated photonic will also play a key role toward space-readiness and aid in the miniaturization of key optical components. However, sustaining the precise stabilization and noise suppression performance of current laboratory setups in these future space-ready solutions will be a challenge.  
%However, a noteworthy caveat in our conclusion here is that our evaluation is based solely on recent experimental findings over terrestrial links. When future ground-to-satellite and satellite-to-satellite optical clock synchronization experiments eventually take place, issues such as Doppler shifts due to satellite motion will need to be accounted. The performance of classical  frequency combs in Doppler shifted environments is another open field of research.

%Over a $113$ km turbulent terrestrial link, O-TWTFT can achieve $10^{-18}$ uncertainty in $10^3$ seconds of integration\cite{Shen2022} which is sufficient performance to support current optical clocks. However, conventional LOS demands high signal power levels at the receiver (of order nanowatts), which would be infeasible in size, weight, and power (SWaP) constrained satellite links. Fortunately, adding a time programmable frequency comb (TPFC) and Kalman filtering to LOS can reduce the power demand to order femtowatts\cite{Caldwell2023}. In recent work, a TPFC-based setup successfully reached an uncertainty of $10^{-18}$ within $10^3$ seconds over a $300$ km terrestrial link. This performance was achieved despite experiencing $70\%$ signal fades which reduced the received signal power to $150$ femtowatts \cite{Caldwell2022,Caldwell2023}. In \cref{fig:figure2}, we show a comparison between the precision limits experimentally achieved by the TPFC lab-based experiment\cite{Caldwell2022}, TPFC over a free-space $300$ km link\cite{Caldwell2023} and GPS\cite{Droste2015}.  

\section{\label{sec:quantum-combs}Quantum frequency combs}
In this section, we provide details on the potential advantage (i.e. quantum advantage) of using a quantum frequency comb instead of a classical frequency comb as the signal in clock synchronization. We show how quantum frequency combs are generated from classical frequency combs, the unique properties that quantum-based techniques exploit, the advantage in theory and the main challenges for a free-space implementation of quantum-enhanced clock synchronization.

\subsection{\label{sec:quantum-efficient-generation}Generating quantum frequency combs}
Let us begin with an outline of common methods for generating quantum frequency combs. The generation stage of a quantum frequency comb strictly requires an optical non-linearity which is commonly a $\chi^{(2)}$ or $\chi^{(3)}$ process. Both nonlinear processes are probabilistic in nature, and are based on the dielectric material used in the optical setup. The probabilistic nature also limits the generation efficiency of quantum frequency combs. In this work, we focus on $\chi^{(2)}$ and its implementation in spontaneous parametric down conversion (SPDC).

\begin{figure}
    \centering
    \includegraphics[width=\linewidth]{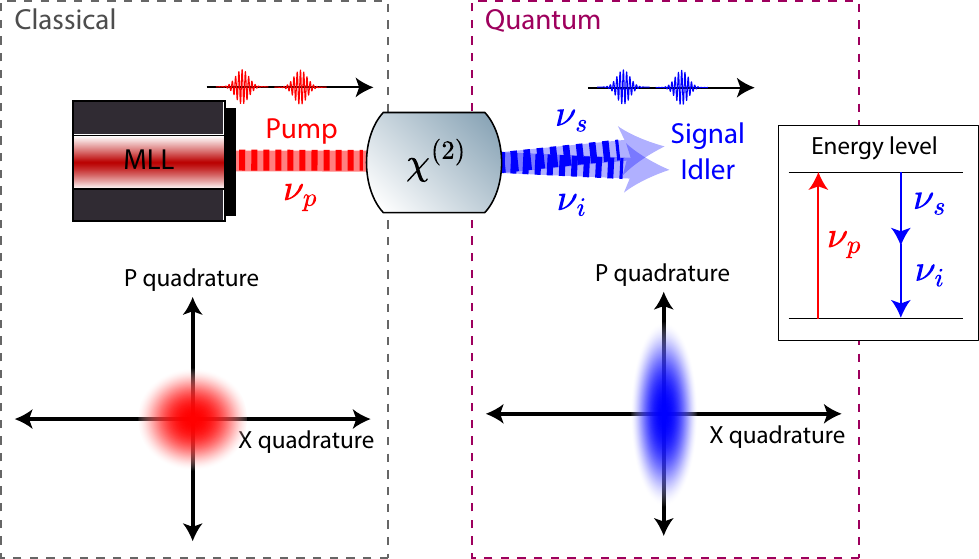}
    \caption{The spontaneous parametric down conversion (SPDC) process is a $\chi^{(2)}$ non-linear process that can be used to convert a classical frequency comb (e.g. mode-locked laser, MLL) into a quantum frequency comb that exhibits quadrature-squeezing properties. The implementation of SPDC shown above is degenerate whereby the signal and idler have exactly half the energy (in turn center frequency) as the pump. As an example, we show the phase space diagram of the classical and quantum frequency comb with X-quadrature-squeezing. Evidently, the quantum frequency comb has reduced variance in the X quadrature and would correspondingly yield a more precise measurement than the classical frequency comb.}% Acronyms and symbols used are as follows: \textbf{MLL}: mode-locked laser, \textbf{SPDC}: spontaneous parametric down-conversion, $\nu_p$: center-frequency of pump, $\nu_s$: center-frequency of signal, and $\nu_i$: center-frequency of idler.}
    \label{fig:squeezed-spdc}
\end{figure}
The $\chi^{(2)}$ non-linearity can be induced using a BIBO crystal 
(i.e. \ch{BiB3O6})\cite{Patera2010a,MedeirosDeAraujo2014,Kobayashi2015,Wang2018,LaVolpe2020}. When an MLL-based classical frequency comb is used as the input pump, the classical frequency comb will convert into a quantum frequency comb after passing through the crystal and undergoing a process known as SPDC. The conversion process can be from a single-pass through the crystal or after multiple passes when the crystal is placed inside a resonant cavity\cite{DeValcarcel2006a,Patera2008,Patera2010a}. In principle, SPDC consists of converting a ``pump'' into a ``signal'' and an ``idler''.

When the pump is a monochromatic CW laser centered at frequency $\nu_p$, the SPDC output is a monochromatic CW signal at frequency $\nu_s$ and a monochromatic CW idler at $\nu_i$ such that,
\begin{gather}
    \nu_p=\nu_s+\nu_i\quad\text{and}\quad\vec{k}_p=\vec{k}_s+\vec{k}_i ,
    \label{eq:spdc-equation}
\end{gather}
where $\vec{k}_{p,s,i}$ are the momentum vectors of the pump, signal and idler respectively. The SPDC is considered degenerate when the signal and idler are indistinguishable in terms of their center frequency, direction of travel and polarization, i.e. ${\nu_s=\nu_i=\nu_p/2}$. Degenerate operation is desired as this induces phase-sensitive variance in the signal and idler quadratures where the variance is reduced below the SQL at certain phase --- a phenomenon known as quadrature-squeezing\cite{Wu1986,Lvovsky2015}. 
However, degenerate SPDC is challenging to achieve in practice and depends strongly on factors such as the pump power and crystal temperature conditions\cite{Andersen2016}. Single-pass SPDC also suffers from low efficiency since a large portion of the pump photons usually pass through the crystal unconverted\cite{Couteau2018}. In \cref{fig:squeezed-spdc}, we have shown how single-pass SPDC is used to produce quadrature-squeezed states.

Let us take a brief aside to define the term ``quadrature''. Quadrature refers to the quantum-version of the real and imaginary components of an electromagnetic wave\cite{BachorHansA}. For instance, an electric-field amplitude, $E(\vec{r},t)$, with spatial coordinate vector $\vec{r}$ and time coordinate $t$ is given by\cite{BachorHansA}
\begin{gather}
    E(\vec{r},t)\propto X(\vec{r},t)\cos(2\pi\nu_0t)+P(\vec{r},t)\sin(2\pi\nu_0t),
\end{gather}
where $X(\vec{r},t)$ and $P(\vec{r},t)$ are the quadrature amplitudes with real numbers. In the quantum perspective, the quadratures are unit-less operators with an expectation and standard deviation, which we denote in this work as simple $X$ and $P$, respectively. A quadrature-squeezed quantum state has a variance in one quadrature that is below the SQL and in the other quadrature the variance is above the SQL\cite{Lvovsky2015}, as shown in \cref{fig:squeezed-spdc}. 

Returning to SPDC, improving the efficiency of single-pass SPDC is crucial for producing a useful quantum frequency comb. Two main strategies in the literature include using a pulsed pump such as an MLL instead of a CW pump, and placing the crystal inside a resonant cavity\cite{Caves1981,Jiang2012a,Pinel2012a,Jiang2012a,Roslund2014,Vahlbruch2016,Fabre2020a,Yang2023}. An MLL has a higher peak power relative to an equivalent CW laser which considerably improves the conversion efficiency. Further, a resonant cavity builds-up of energy with each pass further increasing the peak power. The optical parametric oscillation (OPO) is a parametric oscillator that satisfies both efficiency strategies. A version of the OPO tailored for a pumped laser system is called the synchronously pumped optical parametric oscillator (SPOPO). The SPOPO uses a cavity that has a precisely-controlled length that is locked to the pumping MLL's $f_r$ (and in turn the round-trip time of a single pulse). The state-of-the-art conversion efficiency of a SPOPO system is currently at approximately $20\%$\cite{Gao2024}. For degenerate SPDC within the SPOPO, the MLL pump must have a peak power that is below (but as close as possible to) the oscillation threshold of the cavity. Operating near the cavity threshold also helps to maximize the quadrature-squeezing level produced in the output\cite{Jiang2012a}. A visual diagram of a typical SPOPO setup with a second harmonic generator and an OPO is provided in \cref{fig:lamine}.

An MLL has a broadband spectrum with a large number of frequency modes (in practice around $10^5$ modes\cite{Pinel2012a}) as described by \cref{eq:comb-equation}. Each of these frequency modes will down-convert during SPDC following \cref{eq:spdc-equation}, and produce a considerably more complex signal and idler than in the monochromatic CW case. Particularly, in the degenerate case, both signal and idler will span an identical range of frequency modes $\nu_s=\nu_i=(Nf_r+f_0)/2$ that exhibits multi-partite entanglement between the frequency modes\cite{MedeirosDeAraujo2014}. For the purpose of clock synchronization, we are instead interested in an alternative (simpler) description of the ``complex'' signal and idler state which can be achieved via a unitary change in the basis\cite{MedeirosDeAraujo2014,Patera2010a,DeValcarcel2006a,Pinel2012a}. Specifically, the eigenmodes of the SPOPO, which are constructed by taking a linear combination over all frequency modes, gives a new basis in which each the signal and idler can be characterized by the outgoing pulse spectral amplitude and phase profiles\cite{Pinel2012a}. In the eigenmode basis, quadrature-squeezing is observed when the MLL pump has a peak power below the cavity threshold. Visualizations of the X-quadrature-squeezed quantum frequency comb generated by a SPOPO are provided in \cref{fig:squeezed-spdc,fig:lamine}.  

%In addition, depending on the measurement basis used, the same SPOPO output can be shown to exhibit multipartite entanglement across the frequency modes\cite{Chen2013,MedeirosDeAraujo2014,Gerke2015}. 
\begin{figure*}
    \centering
    \includegraphics[width=\linewidth]{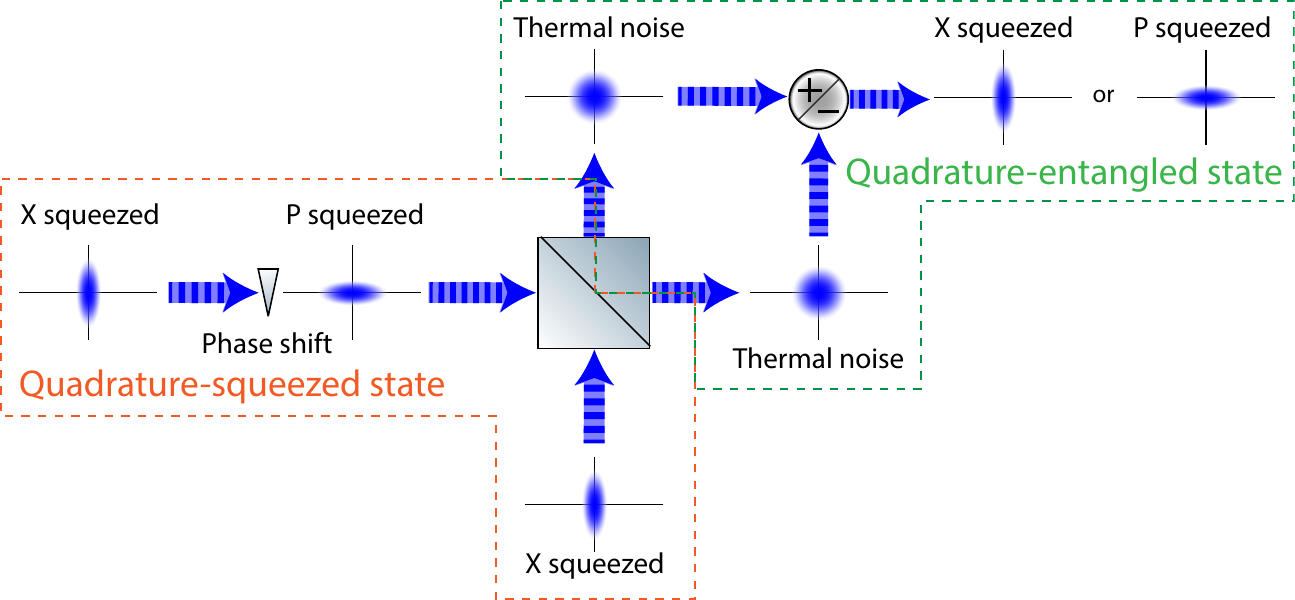}
    \caption{A method for producing a quadrature-entangled quantum frequency comb by mixing two quadrature-squeezed states onto a $50:50$ BS\cite{Eberle2013,Lvovsky2015,Gosalia2023}. The two inputs must be quadrature-squeezed in orthogonal quadratures, and this can be achieved (for example) by using a $\pi/2$ phase shifting stage on one of the input arms. After a sum/difference stage, the final output will exhibit quadrature-squeezing. In a recent study\cite{Gosalia2023}, we compared the performance of a quadrature-entangled state against a quadrature-squeezed state\cite{Gosalia2022} for inter-satellite quantum-enhanced clock synchronization. }
    \label{fig:quadrature-entangled}
\end{figure*}
Quantum frequency combs with quadrature-entanglement can also be produced using the SPOPO method. In this case, two quadrature-squeezed quantum frequency combs would be mixed on a $50:50$ beam splitter (BS). Before the mixing, it will be important to ensure that the two quantum frequency combs exhibit squeezing in different quadratures which can be achieved by applying a path delay in one of the paths equivalent to a $\pi/2$ phase delay. The two output signals from the BS would exhibit variance below the SQL in the sum and difference quadrature modes --- thus the quadratures are entangled. This particular configuration for producing quadrature-entanglement has been implemented\cite{Eberle2013} and is visualized in \cref{fig:quadrature-entangled}.

While SPOPOs have been used widely in many laboratory experiments, key issues exist when we consider extending the technique to space. Most notably, miniaturizing the SPOPO cavity while sustaining high-efficiency and high-bandwidth in the output has proven to be a challenging task to date\cite{Mondain2019,Bruch2019,Zhang2023}. Although, in a recent study by Stokowski et al.\cite{Stokowski2024}, a new technique was developed which demonstrates an operating OPO on a small-form integrated device based on thin-film lithium niobate. The conversion efficiency achieved on this device was approximately $34\%$ which is at the current state-of-the-art. More pathways to achieving OPO on integrated photonic devices are required as the field evolves further, and will be a key road block for quantum frequency combs on satellites. 

Finally we note that the $\chi^{(3)}$ non-linear processes, also called the Kerr effect, is yet another technique for producing quantum frequency combs. The Kerr effect is a non-linear optical process called four-wave mixing (FWM). In FWM, a CW classical laser is converted into a quantum frequency comb. However, since we are most interested in generating quadrature-squeezing and quadrature-entanglement, the FWM output is not currently suitable for our application. Specifically, an FWM-based quantum frequency comb has quadrature-squeezing and -entanglement among individual frequency modes\cite{Chembo2016,Chembo2016a}, rather than in a basis that spans over all frequency modes (as in the SPOPO case). Individual frequency modes are beyond the scope of current quantum-enhanced clock synchronization techniques and, as a result, FWM-based quantum frequency combs are not detailed further here. Nonetheless we refer the reader to a few recent studies for reference\cite{DelHaye2007,Kippenberg2011,Lee2012,Bao2014,Dutt2015,Chembo2016,Chembo2016a,Mayer2016,Yang2021}.

\subsection{\label{sec:quantum-fundamental}Optimal clock synchronization}
\begin{figure*}
    \centering
    \includegraphics[width=\linewidth]{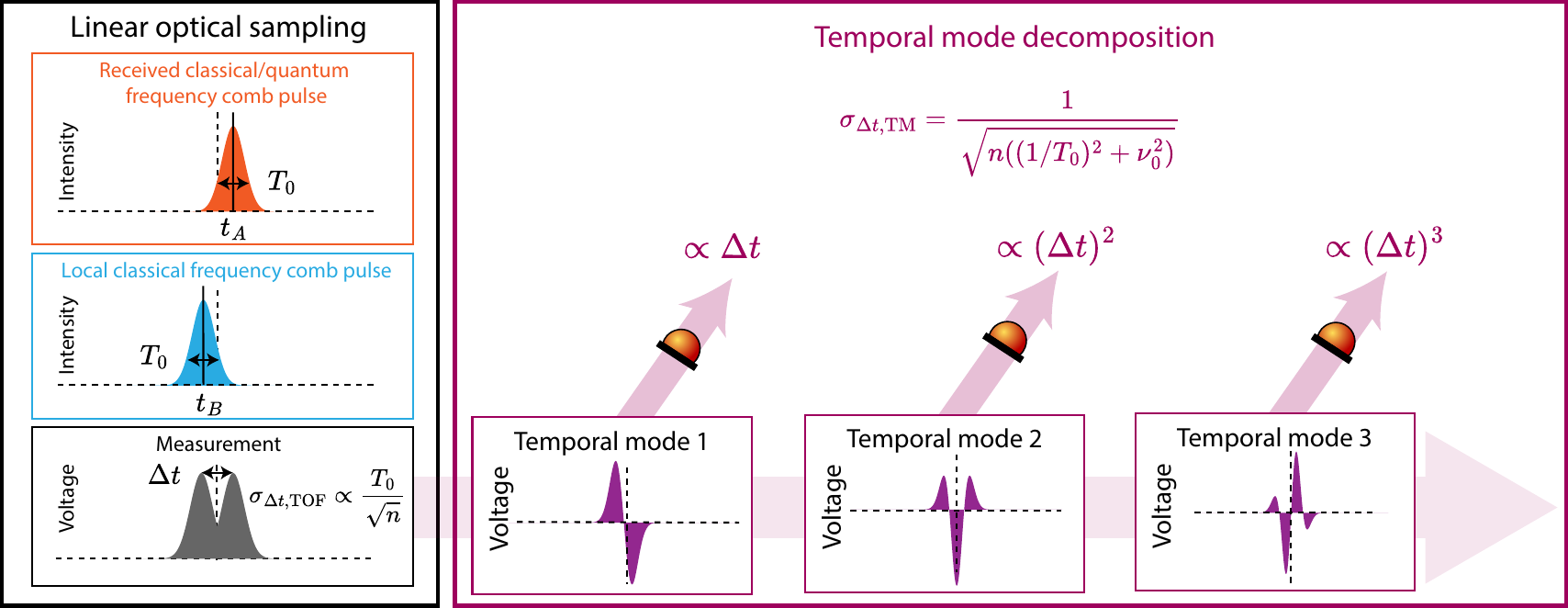}
    \caption{The linear optical sampling (LOS) method as used in O-TWTFT consists of estimating the timing offset, $\Delta t$, between two classical frequency comb pulses (i.e. ${\Delta t=t_A-t_B}$) using interferometry. On the right we show an optimal measurement strategy --- temporal mode decomposition, where the LOS measurement is projected onto higher-order temporal modes before the temporal modes are separately measured. The photocurrent outputs of each temporal mode measurement $\ell$ is proportional to $(\Delta t)^{\ell}$ with a standard deviation of $\Delta t$ measurements given by \cref{eq:tm}. Substituting the classical frequency comb with a quantum frequency comb as the signal would yield a standard deviation scaling that approaches the HL given by \cref{eq:tm-quantum}.}% Acronyms used are as follows: \textbf{LOS}: linear optical sampling, \textbf{O-TWTFT}: optical two-way time and frequency transfer, \textbf{SQL}: standard quantum limit, and \textbf{HL}: Heisenberg limit.}
    \label{fig:LOS}
\end{figure*}
The main objective of quantum-enhanced clock synchronization is to extend the performance of classical clock synchronization from the SQL toward the HL. A quantum frequency comb is one key tool to this end, but, in addition, we also need an optimal measurement strategy. Although the LOS technique is currently a popular method for estimating the timing offset (i.e. $\Delta t$ in \cref{fig:LOS}) between two classical frequency combs, it is not the optimal measurement strategy according to quantum estimation theory\cite{Lamine2008,Jiang2012a}. One example of an optimal measurement strategy for estimating $\Delta t$ is a technique we refer to here as ``temporal mode decomposition''. In temporal mode decomposition, the combined signal from the received (classical or quantum) frequency comb and local classical frequency comb is projected onto higher-order orthogonal temporal modes and the intensity of each temporal mode is measured separately, as shown in \cref{fig:LOS} and discussed in previous works\cite{Lamine2008,Jiang2012a,Ansari2018,Donohue2018,Ansari2021}. A classical frequency comb signal with temporal mode decomposition can reach the SQL\cite{Lamine2008}. A quantum frequency comb signal with temporal mode decomposition will exceed the SQL and approach the HL. %To aid with distinguishing between different measurement strategies, we can compare the ultimate precision limits of each. Let us first begin by detailing the most important measurement parameters in clock synchronization.

\subsubsection{Reaching the SQL with temporal mode decomposition}
In an experiment implementing clock synchronization, several samples of $\Delta t$ are collected to calculate the average timing offset, $\overline{\Delta t}$, and corresponding standard deviation, $\sigma_{\Delta t}$. Note, $\sigma_{\Delta t}$ and TDEV are related, where the former is a statistical standard deviation of $\Delta t$, the latter is the average two-sample deviation of $\Delta t$ (discussed in \cref{sec:ffi-tdev}). Ideally we want $\sigma_{\Delta t}\rightarrow0$ by averaging over a large number of samples of $\Delta t$, however experimental challenges such as restrictions to the transmit power, channel-based losses, and detection inefficiencies limit the number of samples that can be collected in practice. Also, different clock synchronization strategies can achieve different $\sigma_{\Delta t}$ over a finite number of samples over the same link. Hence, finding a measurement strategy that minimizes $\sigma_{\Delta t}$ (i.e. an optimal measurement strategy) is desired.

Measurement strategies that use TOF, such as LOS, have a measurement standard deviation that scales with the total number of photons collected ($n$) given by\cite{Lamine2008}
\begin{gather}
    \sigma_{\Delta t,\text{TOF}} \propto \frac{T_0}{\sqrt{n}}.
    \label{eq:tof}
\end{gather}
Here, minimizing $T_0$ (with ultra-short duration pulses) and maximizing $n$ are the only two strategies for minimizing the standard deviation.
Alternatively, measurement strategies based on the phase difference between a remote and local signal, $\Delta \phi$, (which is in turn related to $\Delta t:=\Delta\phi/(2\pi\nu_0)$) include OFT. The phase method has a standard deviation that scales as\cite{Lamine2008}
\begin{gather}
    \sigma_{\Delta t,\text{phase}} \propto \frac{1}{\nu_0\sqrt{n}},
    \label{eq:ph}
\end{gather}
where $\nu_0$, recall, is the nominal frequency of the signal. Finally, the temporal mode decomposition method provides a superior scaling of the standard deviation compared to both the TOF and phase measurements, which is given by\cite{Lamine2008,Jiang2012a}
\begin{gather}
    \sigma_{\Delta t,\text{TM}} \propto \frac{1}{\sqrt{n\left((1/T_0)^2+\nu_0^2\right)}} ,
    \label{eq:tm}
\end{gather}
where TM denotes temporal modes. The difference in the standard deviation here is due to the fact that the temporal modes of a quantum frequency comb pulse carry both the TOF and the phase information in their temporal profiles. Additionally, temporal mode decomposition has also been shown to be an \textit{optimal} measurement strategy according to quantum estimation theory\cite{Jiang2012a} and therefore \cref{eq:tm} represents the lowest bound in the standard deviation achievable by any measurement strategy when a classical frequency comb signal is used. %Note also that the classical frequency comb (centered at optical frequencies) has both a short $T_0$ and a large $\nu_0$, and therefore would have a lower $\sigma_{\Delta t}$ than other types of classical signals such as CW laser. 

\subsubsection{Reaching the HL with a quantum frequency comb}
There remains one final strategy that can be used to lower $\sigma_{\Delta t}$ further and that is to substitute the classical frequency comb signal with a quantum frequency comb signal. A quantum frequency comb with quadrature-squeezing of factor $r$ would retrieve a standard deviation that scales as\cite{Lamine2008}
\begin{gather}
    \sigma_{\Delta t,\text{TM}}\propto\frac{\exp(-r)}{\sqrt{n\left((1/T_0)^2+\nu_0^2\right)}}\rightarrow\frac{1}{n\sqrt{(1/T_0)^2+\nu_0^2}},
    \label{eq:tm-quantum}
\end{gather}
where the right-hand side is reached as $r\rightarrow\infty$, see\footnote{In a quadrature-squeezed state\cite{Lamine2008}, ${n=\sinh^2(r)=\exp(-2r)+\exp(2r)-1/2}$. As ${r\rightarrow\infty}$, ${n\rightarrow\exp(2r)}$. Therefore, $\sigma_{\Delta t,\text{TM}}\propto\exp(-r)/(n[(1/T_0)^2+\nu_0^2]^{1/2})\rightarrow\exp(-r)/(\exp(r)[(1/T_0)^2+\nu_0^2]^{1/2})=1/(n[(1/T_0)^2+\nu_0^2]^{1/2})$} for specific details on the derivation. Therefore, using a quantum frequency comb and temporal mode decomposition allows us to reach the HL, i.e. ${\sigma_{\Delta t}\propto1/n}$.

\begin{figure*}
    \centering
    \includegraphics[width=\linewidth]{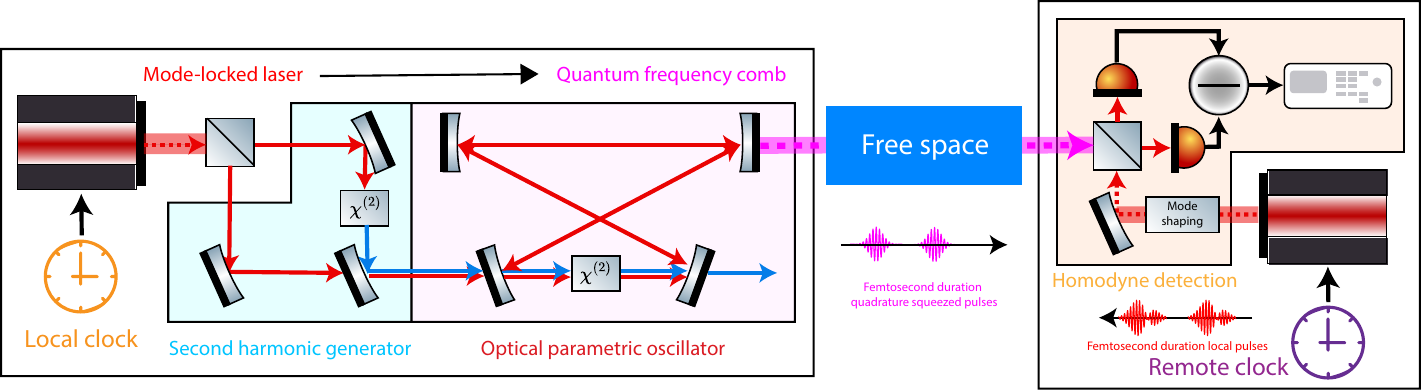}
    \caption{A potential setup for quantum-enhanced clock synchronization using the scheme proposed by Lamine et al.\cite{Lamine2008}. An MLL phase-locked to a local optical clock is quadrature-squeezed via a second harmonic generator and optical parametric oscillator setup before being transmitted over free-space to a remote site. At the remote site, a second MLL is temporally-shaped to a specific superposition of higher-order temporal mode before mixing with the incoming signal. BHD produces an electronic photocurrent that is proportional to the temporal offset, $\Delta t$, between local and remote frequency comb pulses (and in turn the difference between the local and remote clocks).}
    \label{fig:lamine}
\end{figure*}
Initial studies by Giovanneti et al.\cite{Giovannetti2001,Giovannetti2002,Giovannetti2004,Giovannetti2011} inspired research into quantum-enhanced clock synchronization. Recently, a proposal by Lamine et al.\cite{Lamine2008} discusses a practical pathway toward achieving the HL using a quantum frequency comb and a balanced homodyne detection (BHD) setup. In the proposal by Lamine et al.\cite{Lamine2008}, a quadrature-squeezed quantum frequency comb is recommended as the signal that is phase-locked to a local clock. A second classical frequency comb is recommended at a remote site which is phase-locked to a remote clock. The remote classical frequency comb also needs to have a temporal profile that is carefully shaped into higher-order temporal modes using, for instance, a pulse shaper and a spatial light modulator\cite{Patera2008,Fabre2020a}, so that the incoming quantum frequency comb can be projected onto higher-order temporal modes to conduct temporal mode decomposition. 

In \cref{fig:lamine} we show a diagram of a free-space experimental setup for the proposed scheme by Lamine et al.\cite{Lamine2008} which we have investigated in our previous works\cite{Gosalia2022,Gosalia2023}.
The local quantum frequency comb is transmitted over free-space to the remote site, where the two frequency combs are mixed on a $50:50$ BS that is part of the BHD setup. The BHD setup projects the incoming signal onto the temporal modes of the remote classical frequency comb and extract an electrical signal that is proportional to $\Delta t$. After several samples are collected, $\sigma_{\Delta t}$ will scale as per \cref{eq:tm-quantum}. %Note, if the local site used a classical frequency comb instead of a quantum frequency comb, the final uncertainty would scale as \cref{eq:tm}.

\begin{figure*}
    \centering
    \includegraphics[width=\linewidth]{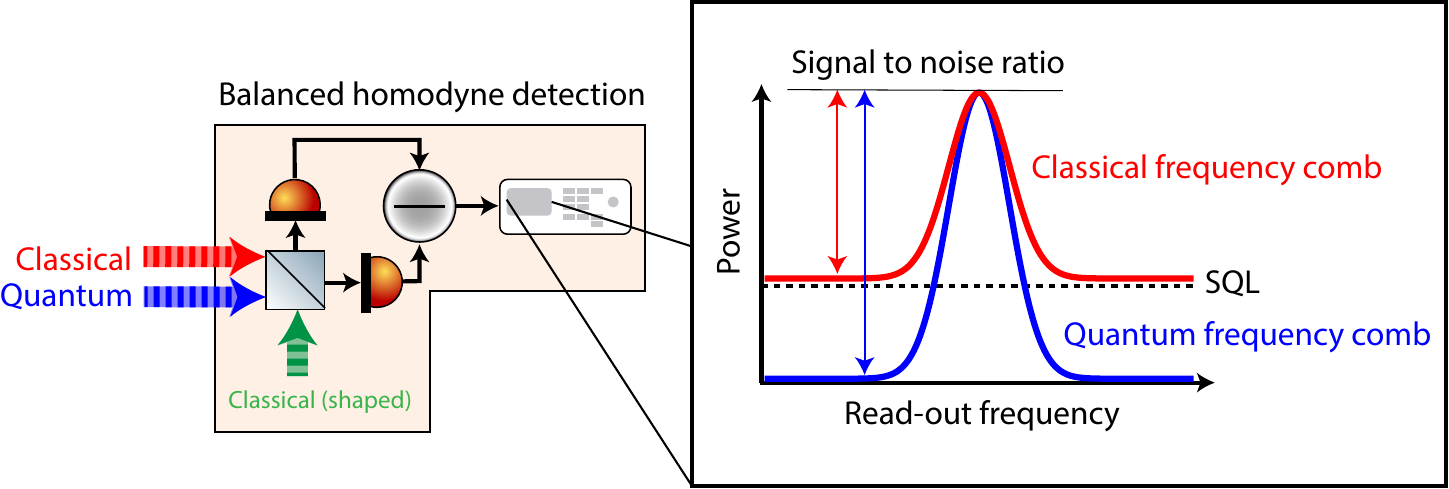}
    \caption{The expected advantage of using a quantum frequency comb relative to a classical frequency comb in clock synchronization over a BHD setup. In red is the classical frequency comb which has a noise floor at the SQL. In blue is the quantum frequency comb that would have a reduced noise floor proportional to the level of quadrature-squeezing. Hence, the quantum approach shows a pathway to achieving a higher signal-to-noise ratio ratio by reducing the noise floor. Note that quantum advantage has been experimentally demonstrated by Wang et al.\cite{Wang2018}.}
    \label{fig:quantum-vs-classical-bhd}
\end{figure*}
In a laboratory experiment, Wang et al.\cite{Wang2018} successfully conducted temporal mode decomposition to achieve a $\sigma_{\Delta t,\text{TM}}$ as per \cref{eq:tm-quantum} with a $1.5$~dB quadrature-squeezed quantum frequency comb. The setup consisted of a Ti:Sapphire-based MLL that generated (ultra-short) $130$~femtosecond duration pulses with ${\nu_0=0.25}$~THz and ${f_r=75}$~MHz. The remote classical frequency comb was produced using the same MLL source. The $1.5$~dB of quadrature-squeezing was achieved using an SPOPO setup\cite{Jiang2012a,Pinel2012a,Kobayashi2015a} which successfully reduced the $\sigma_{\Delta t}$ from ${8.9\times10^{-23}}$~s to ${7.5\times10^{-23}}$~s. These laboratory results confirm that temporal mode decomposition can be conducted in practice to yield quantum advantage. To aid in gaining an intuitive understanding of the result, in \cref{fig:quantum-vs-classical-bhd} we show the difference between the expected measurement result using a quantum vs. a classical frequency comb as signal with temporal mode decomposition. Based on the signal-to-noise ratio ratio, it can deduced that the quantum-approach would be able to resolve weaker signal levels and thereby provide precision enhancements for clock synchronization.

\subsection{Challenges to achieving quantum advantage in practice}
%An implicit assumption made so far in achieving \cref{eq:fundamental} is that $\omega_0$ and $\Delta\omega$ are constant during clock synchronization. However, as discussed in \cref{sec:classical}, stabilizing a frequency comb requires monitoring and control equipment. For both classical and quantum frequency comb-based clock synchronization, mode-locking and noise elimination are essential to reach the SQL performance. % $\omega_0$ and $\Delta\omega$ is challenging in practice, and is a performance-limiting factor for the widespread use of frequency combs in many applications including clock synchronization. An approach gaining momentum for generating highly stable frequency combs is mode-locked lasers (MLL). MLLs coherently lock multiple modes of light in a laser cavity and produce pulses that progressively shorten in time and widen in frequency spectrum with every repetition of a cavity roundtrip\cite{Haus2000, Fortier2019,Diddams2020}. %The working principles of a frequency comb are summarized in \cref{fig:freq-comb}.
%We define a quantum advantage as the conditions that allow a quantum frequency comb to yield a genuine precision-to-resource advantage compared to an equivalent classical frequency comb. 
The results by Wang et al.\cite{Wang2018} have inspired our recent work on satellite-based quantum-enhanced clock synchronization in LEO\cite{Gosalia2022,Gosalia2023}. Our work has revealed some opportunities as well as challenges for the temporal mode decomposition method and, in turn, achieving quantum advantage in practice. Particularly, we have found that that quantum advantage is highly sensitive to beam diffraction, satellite pointing misalignment and photodetection inefficiencies.
In addition, differences between the generation efficiency of classical vs. quantum frequency combs impact the level of $n$ captured in practice and, in turn, the $\sigma_{\Delta t}$ achievable. The practical realization of quantum advantage thereby relies on three important factors: (1) the efficient generation of quantum frequency combs, (2) the minimal impact of noise and loss during free-space propagation, and (3) the near-perfect detection of the quantum frequency comb at the remote site. 

Precision beyond the SQL, towards the HL, relies strongly on the ability to generate \textit{useful} quantum properties of light. Unfortunately, current state-of-the-art laboratory methods for generating quantum frequency combs have low efficiencies --- in some state-of-the-art laboratory-setups, the overall conversion efficiency is as low as $10\%$\cite{Dong2023}. On the other hand, state-of-the-art studies with classical frequency combs have demonstrated conversion efficiencies reaching $50\%$ and better\cite{Chang2022}. Based on this difference alone, for $n$ quantum frequency comb photons vs. $n$ classical frequency comb photons, a $>5\times$ reduction of $\sigma_{\Delta t}$ would be required at a minimum to justify deploying the quantum approach. In terms of the equivalent quadrature-squeezing level, this equates to a requirement of $>7$~dB of quadrature-squeezing\footnote{Using the formula $-10\log_{10}\left(1/5\right)\approx7$ dB, where it is assumed that the standard quantum limit noise is normalized to $1$.}, which is near the state-of-the-art. For this reason, in coming years more research focus is required in improving the efficiency of generating quadrature-squeezed quantum frequency combs.%More research focus in the efficient generation of quantum frequency combs is needed.

Quadrature-squeezing and quadrature-entanglement properties are also highly sensitive to photonic loss and noise issues. In our work\cite{Gosalia2022,Gosalia2023}, we found that beam diffraction and satellite pointing issues contribute as loss and noise, respectively, to a quantum frequency comb\cite{Gosalia2022,Gosalia2023}. Whilst the noise contributions due to satellite pointing could be controlled by using fine-tracking control systems for real-time tracking of satellites and ground stations on-board\cite{Gosalia2023a,Song2017,Zhang2020,Madni2021}, photonic loss over the channel due to beam diffraction and turbulence is unfortunately irrecoverable. To compensate for the consequent degradation of quantum properties, we could use ``stronger'' quantum properties (i.e. higher squeezing level), or sophisticated loss compensation techniques\cite{Zhou2018,Shettell2021} with noiseless amplification\cite{Frascella2021} and degenerate parametric amplification\cite{Shaked2018,Knyazev2019,Frascella2021} at the remote receiver. These techniques have not yet been studied in the context of quantum-enhanced clock synchronization, and are open directions of much-needed research.%Determining a manner to incorporate these solutions while still minimizing the SWaP [??? i need a check on all grammar around this acronym, often incorrect, read aloud the words to see] is yet another under-studied research direction needing attention.% In \cref{sec:quantum-free-space}, we further discuss the issues of transferring quantum frequency combs over free-space.

Finally, the photodetectors (inside the BHD setup) can also contribute loss and noise to the quantum frequency comb if they have a photodetection efficiency less than $100\%$\cite{Gosalia2022}. Fortunately, state-of-the-art photodetection efficiencies, achieved in laboratory settings, have been of order $90\%$ and greater\cite{Zhou2018,Wang2019}, with one recent experiment even reaching $99.5\%$ efficiency\cite{Vahlbruch2016}. In the near future, we expect similar levels of performance on-board satellites --- however, achieving $90\%$ photodetection efficiency on a chip-based platforms (with low SWaP demand) is yet another on-going challenge in the field\cite{Yue2018}. Recent work on super-conducting nanowire-based balanced homodyne detection setups have demonstrated successful operation with $77\%$ photodetection efficiency in a low size and weight packaging\cite{Protte2024}. Although, super-conducting solutions require cryogenics which would add to the overall SWaP demand. Implementation of BHD in space is yet another open field of research.%Although, since only a few passive optical components are required (i.e., a beam splitter and two photodetectors) the setup should, in theory, have low SWaP demand.
%In our previous study\cite{Gosalia2022}, we predicted that a free-space inter-satellite link up to $100$~km could achieve a quantum advantage with $15$~dB of squeezing and an aperture of radius $60$~cm. However, we found that satellite pointing jitter considerably reduces the inter-satellite range range considerably, with $1$~$\mu$rad of jitter causing up to $30$~km reduction. In another study\cite{Gosalia2023}, we considered a quadrature-entangled instead of a quadrature-squeezed signal and showed that a two-mode signal could provide a form of diversity for overcoming channel loss effects. 
%In summary, a quantum frequency comb, in principle, should provide a precision-to-resource advantage over a classical frequency comb when various non-ideal factors are mitigated. However, a practical realization of this quantum advantage, especially over satellite links, requires close attention to the generation, transmission, and measurement stages. All three of these stages should have minimal loss and noise impact on the final quantum signal, otherwise the quantum advantage will be at risk. %Therefore an non-trivial trade-off between classical and quantum frequency combs exists in practice, which up till now has been in the favor of the classical approach, due to setup simplicity, but may change in the future with research maturity.

In our recent study, we found that at least $15$~dB of quadrature-squeezing is required to achieve $2\times$ quantum advantage over a typical $100$~km inter-satellite link that consists of LEO satellites with telescopic apertures of $0.3$~m radius ($0.6$~m diameter) and photodetectors operating at $90\%$ efficiency\cite{Gosalia2022}. Unfortunately, $15$~dB quadrature-squeezing is at the current state-of-the-art\cite{Vahlbruch2016}, and even higher squeezing levels would dramatically increase the SWaP demand of the system. Additionally, we found that the satellite pointing angle jitter of both the local and the remote satellites needs to be within a standard deviation of 1~$\mu$rad. If the standard deviation is above this level, the consequential degradation would eliminate any quantum advantage. Hence, a significant engineering effort is required for inter-satellite transfer of quadrature-squeezed quantum frequency combs. Additionally, we found that photonic loss due to beam diffraction limits the range of the inter-satellite links to within $300$~km. For longer-range inter-satellite links, unfortunately multiple LEO satellites would be needed in a ``multi-hop'' configuration. 

\subsection{\label{sec:quantum-free-space}Quadrature-squeezing vs. quadrature-entanglement over free-space}
%During clock synchronization, timing information is exchanged via pulses of light. When a quantum frequency comb is used as the source of the pulses, the survivability of the quantum properties during the exchange becomes paramount. 
Although quadrature-squeezing has largely been the focus of our discussion, quadrature-entanglement could, in principle, be used instead to yield a squeezed variance as discussed in \cref{sec:quantum-efficient-generation} and shown in \cref{fig:quadrature-entangled}. Although, our study on quadrature-squeezing has revealed certain system characteristics that are required in order to achieving quantum advantage over inter-satellite links\cite{Gosalia2022}. %An open research question includes how to overcome loss and noise issues on squeezed light as, without any mitigation strategies, current progress in this field is limited by the fragility of squeezing. %Moreover it is noteworthy that the generation efficiency was not accounted in these preliminary calculations.
Specifically, quadrature-entanglement provides an alternative quantum property that could be exploited to achieve the HL. However, it is important to note that, in principle, the scaling of $\sigma_{\Delta t}$ using a quadrature-entangled quantum frequency comb is the exact same as quadrature-squeezing, as shown by us in a previous work\cite{Gosalia2023}. In other words, the resulting signal-to-noise ratio ratio gain achieved by using a quadrature-entangled quantum frequency comb would be the same as a quadrature-squeezed comb of equivalent $r$. Instead, the advantage of a quadrature-entanglement is in the greater resilience this state provides over lossy free-space channels. Particularly, the quadrature-correlations shared between pairs of quadrature-entangled quantum frequency comb pulses can be taken advantage to recover over asymmetrically lossy channels\cite{Gosalia2023}. Asymmetric loss is expected over dynamically varying channels such as inter-satellite and ground-satellite links, hence in these contexts, quadrature-entanglement may be a better solution than quadrature-squeezing. Although, much more rigirious study both in theory and experiments are needed. 
%Interestingly, we note that since a quantum frequency comb can exhibit both squeezing and entanglement properties, depending on the mode basis in which it is measured\cite{MedeirosDeAraujo2014}, both properties could in principle be used based on the channel conditions. Particularly, as discussed in detail by Meiros de Ara\'ujo et al.\cite{MedeirosDeAraujo2014}, quantum frequency combs exhibit multi-partite entanglement in the frequency mode basis, and quadrature-squeezing in a basis over constructed ["over constructed ??? what does this mean?] by taking a superposition of frequency modes over the entire spectrum spanned by the quantum frequency comb. %However, much alike the classical case, the limiting factor for space-readiness of quantum frequency combs is low SWaP generation, transmission and detection. Toward this end, integrated photonic are providing a potential pathway.

Finally we mention that no previous experimental studies have been conducted on the propagation of quantum frequency combs over turbulent atmospheric channels. Hence, little is known about the impact of turbulence on quadrature-squeezing and quadrature-entanglement over these types of free-space channels. 
%This may suggest that the experimental challenges are currently insurmountable. 
In 2014, Peuntinger et al.\cite{Peuntinger2014} investigated the transfer of polarization-squeezing using a CW laser (not from a quantum frequency comb) over a $1.6$~km free-space channel. Their study found that the polarization degree of freedom offers some level of resilience against the turbulence-induced phase and waveform distortions. This insight may transfer to future experiments with quantum frequency combs.
%The Micius satellite has  shown promising results for discrete-variable-based quantum states of light (e.g. single photons) where quantum teleportation (among many other achievements) was successfully deployed over a $1200$~km link between Micius and two ground stations\cite{Li2022}. These  pioneering  experiments clearly showed that space-based quantum links are indeed viable. However, we note quadrature-squeezing and quadrature-entanglement are  continuous-variable properties not directly related to the Micius experiments.  The performance of quantum frequency combs over similar channels used by Micius and how their superior timing outcomes impact quantum communication over such channels remains an exciting open area of research. 

%\subsection{Recent progress}
%The key distinction between the process for generating a classical and a quantum frequency comb is the pump power used relative to the cavity oscillation threshold power. When the pump power is below or near the threshold, quantum properties such as quadrature-squeezing in the supermodes and multipartite entanglement between frequency modes exist\cite{Patera2008,Jiang2012a,Pinel2012a,Kobayashi2015a,Yang2021}.

%Recently, there has been progress towards on-chip generation of a quantum frequency comb using a microresonator-based approach\cite{Yang2021}. 

%\subsection{Other promising applications for quantum frequency combs}

\section{\label{sec:perspective}Perspective and outlook}
Both classical and quantum frequency combs can be used for satellite-based clock synchronization of a network of optical clocks. MLLs are a highly-stable source of classical frequency combs and it has already been demonstrated, within laboratory settings, that state-of-the-art MLL stabilization and noise suppression levels exceed the requirements to synchronize optical clocks. Further, recent experiments with MLLs in space have provided valuable insights into some of the key engineering strategies required to develop MLLs that can withstand launch acceleration and radiation. Also, developments in the field of integrated photonics will enable chip-based MLL generation in coming years. MLLs can synchronize distant clocks over a $300$~km terrestrial link with O-TWTFT and achieve an FFI and TDEV near the SQL. On the other hand, quantum frequency combs, generated by converting an MLL-based classical frequency comb through a SPOPO, can also exhibit the superior stabilization and noise suppression performance of MLLs with added quantum properties. By combining an optimal measurement strategy (e.g. temporal mode decomposition) with a quantum frequency comb signal, the HL can approached. Since satellites have strict SWaP restrictions, physically realizing performance at the HL will enable much-needed resource efficiencies.

Classical frequency combs are more mature in terms of technical readiness than quantum frequency combs. In our perspective, O-TWTFT could be deployed over satellite-ground and inter-satellite links in the near future. However, future implementations of the LOS method should include temporal mode decomposition for optimal measurement. This step will maximize the precision achievable by LOS and provide robustness against free-space issues such as atmospheric turbulence and Doppler shifts during satellite motion. Although, further research is required to better understand the performance of the temporal mode decomposition method in these non-ideal environments. It is also our perspective that T2L2 could be improved with temporal mode decomposition and BHD with ``slow'' photodetector sampling rates in the microwave region. There may also be scenarios such as on-board low-powered platforms including cubesats, where T2L2 is preferred over O-TWTFT due to less ``complex'' hardware.

The key road-block for quantum frequency combs is the relatively low efficiency with which they can be generated at present. The generation efficiency needs to be improved to a level that is comparable to the generation efficiency of classical frequency combs. Additionally, we want to encourage more transparency on the conversion efficiency values for various quantum frequency comb-generation methods proposed in the literature. A second limitation on quantum frequency combs is the sensitivity to photonic losses and noise. We recommend research effort toward the development of mitigation strategies at the receiver (or transmitter) for enabling the free-space propagation of quadrature-squeezing and quadrature-entanglement over realistic link distances. %Without efficient mitigation strategies, quantum frequency combs may be limited to fiber-based links in practice.   

As a final note, we also mention that there are many fields of research that would benefit from performance near the HL. These applications include spectroscopy\cite{Belsley2022,Picque2019,Shi2023a}, chemistry\cite{Russell2023}, gravitational wave detection\cite{Kwan,Wu2022b}, communications\cite{Corcoran2020}, ranging\cite{Caldwell2022}, and angle estimation\cite{Shimizu2020}. As an illustrative example we mention that, in a recent experiment, a classical frequency comb was used to estimate range down to an uncertainty in distance of $10$~nanometer using only $10$~nanowatts of received power and over a ${\tau=40~\mu\text{s}}$ window\cite{Caldwell2022}. Had a quantum frequency comb been used instead, the uncertainty in distance could have been as low as $4$~picometer (i.e. 4 orders of magnitude smaller) for the same received power level and $\tau$\footnote{Following\cite{Caldwell2022}, the ranging precision, $\sigma_R=\Delta R/(2\ln{2}\sqrt{\eta n_s})$ where $C=1$, $\Delta R=52.6\mu$m, $\eta=1$ and $n_s=10^8$ in our example. The ranging uncertainty for the quantum case would ideally scale as $\sigma_{R,Q}=\Delta R/(2\ln{2}\sqrt{\eta}n_s)$.}. Although, this quantum advantage would be in the most ideal case with infinite squeezing, any finite levels of squeezing can produce orders of magnitude enhancements providing between $1-4$ orders of magnitudes overall. Similar gains could also be expected in future space-borne gravitational wave detectors where a reduction in the laser phase noise below the SQL during arm locking could substantially improve the sensitivity of the apparatus\cite{Wu2022b}. %Hence, the future of quantum frequency combs is bright when the implementation challenges faced today are overcome. 
%Satellite-based challenges...beyond SWaP, Doppler, radiation induced noise
%\textit{\textcolor{blue}{Overall perspective on the future directions of classical and quantum frequency combs for clock synchronization --- main challenges and solutions towards space-readiness. A comparison of quantum frequency combs with other quantum clock synchronization approaches such as quantum clock synchronization (using single photos) and entanglement.}}

\section{\label{sec:conclusion}Conclusion}
%In summary, frequency combs can synchronize a network of optical clocks over free-space. Classical frequency combs can already reach the SQL in precision, and  quantum frequency combs will be required to go beyond this limit. 
In this perspective on the field of satellite-based synchronization, we have reviewed the current state-of-the-art in timing and synchronization, emphasizing the need for an optical-frequency based approach. Further, we have outlined the performance characteristics of various optical schemes based on recent experiments conducted over terrestrial links, highlighting the key advantages offered by a frequency-comb based scheme. 

We have described how the classical frequency comb is robust, highly stable, and a high-precision solution with certain SWaP advantages over other classical methods. Recent experiments with classical frequency combs have shown successful operation over turbulent channels, while satellite-based experiments have sustained mode-locking capabilities in space over year-long missions. Although inter-satellite, ground-to-satellite and satellite-to-ground tests are yet to be conducted, the current state-of-the-art shows promise for successful operation in these settings as well. However, new challenges do lie ahead, including overcoming the effect of Doppler shifts during satellite motion.

We have also described how a quantum frequency comb has the potential to provide precision-to-resource advantages that go beyond the capabilities of a classical frequency comb. 
%In resource-constrained environments like satellite networks, we have outlined how the quantum approach would be more favorable in theory.
In resource-constrained environments such as satellite networks, we have outlined how the quantum approach would, in principle, be a better solution than the equivalent classical approach due to the ability to reach the HL during clock synchronization. 
However, some limiting factors exist with regard to practical implementations, including the generation efficiency and the fragility of squeezing and entanglement over free-space. 
%as evidenced through some of our recent work\cite{Gosalia2022,Gosalia2023,Gosalia2023a}. 
%Furthermore, the quantum approach is not immune to the loss and noise issues encountered by a classical frequency comb (e.g. diffraction, beam jitter, Doppler and more), and thus will likely only be implemented over links where these adverse effects can be contained and/or overcome.

There are many open questions in relation to the use of classical and quantum frequency combs for satellite-based clock synchronization. In terms of technological readiness, we have discussed how classical frequency combs have reached higher maturity and, as such, future research efforts on their use should focus on reducing the SWaP footprint and conducting experiments over satellite-based links on the path to commercialization. With regards to quantum frequency combs; we have highlighted higher-efficiency generation of quantum states is a key road-block at present to further progress, and that better understanding of the mechanisms for overcoming the effects of loss and noise over free-space links are required.
%--- without which the quantum approach, despite its advantages, may be abandoned. 

\section*{Acknowledgments}
R.K.G. is supported by an Australian Government Research Training Program Scholarship and the Sydney Quantum Academy. Approved for Public Release: NG24-1117.

\section*{Author Declarations}
\subsection*{Conflict of Interest}
The authors have no conflicts to disclose.

\subsection*{Author Contributions}
R.K.G. led the conception of the manuscript, wrote the structure, original draft,  and final manuscript. R.M. reviewed the structure, original draft and final manuscript, and supervised the project. H.L., R.A., J.G. and P.B. reviewed the structure, original draft and final manuscript. All authors read and
approved the final manuscript.
%\textbf{Ronakraj K. Gosalia:} Conceptualization (lead); Writing --- original draft; Writing --- review \& editing. \textbf{Robert Malaney:} Conceptualization; Review; Supervision. \textbf{Ryan Aguinaldo:} Conceptualization; Review. \textbf{Jonathan Green:} Conceptualization; Review; \textbf{Holly Leopardi:} Conceptualization; Review. \textbf{Peter Brereton:} Conceptualization; Review.

\section*{Data Availability}
The data that support the findings of this study are available from the corresponding author upon reasonable request.

%\section*{References}
\bibliography{main}{}
\bibliographystyle{plain}

\end{document}